\documentclass[lettersize,journal]{IEEEtran}

\usepackage{amsmath,amsfonts}

\usepackage{array}
\usepackage[caption=false,font=normalsize,labelfont=sf,textfont=sf]{subfig}
\usepackage{textcomp}
\usepackage{stfloats}
\usepackage{url}
\usepackage{verbatim}
\usepackage{graphicx}
\def\BibTeX{{\rm B\kern-.05em{\sc i\kern-.025em b}\kern-.08em
    T\kern-.1667em\lower.7ex\hbox{E}\kern-.125emX}}
\usepackage{balance}
\usepackage{amssymb}
\usepackage{algorithm}
\usepackage{algpseudocode} 
\usepackage[nameinlink, capitalize]{cleveref}
\usepackage{booktabs}
\usepackage{multirow}   
\usepackage{array}      
\usepackage{makecell}   
\usepackage{diagbox}
\Crefname{algorithm}{Algorithm}{Algorithms} 
\allowdisplaybreaks[4]

\DeclareMathOperator{\Diag}{Diag}
\DeclareMathOperator{\blkdiag}{blkdiag}

\begin{document}
\title{Channel Estimation for Beyond Diagonal \\ RIS-Aided Multi-User mmWave Systems}
\author{Linyu~Peng, Tian~Qiu, Cunhua~Pan,~\IEEEmembership{Senior Member,~IEEE}, Jiangzhou~Wang,~\IEEEmembership{Fellow,~IEEE}, Taihao~Zhang,
    and~Hong~Ren,~\IEEEmembership{Member,~IEEE}
\thanks{Linyu Peng, Tian Qiu, Cunhua Pan, Jiangzhou Wang, Taihao Zhang and Hong Ren are with National Mobile Communications Research Laboratory, Southeast University, Nanjing, China (e-mail:213223793, tianqiu, cpan, j.z.wang, taihao, hren@seu.edu.cn).}
% <-this % stops a space
}
\maketitle

\begin{abstract}
Beyond diagonal reconfigurable intelligent surface (BD-RIS) represents a promising architecture for advancing millimeter-wave (mmWave) communications. However, its intricate inter-element connections invalidate the conventional decoupled mathematical structure, thereby severely complicating cascaded channel estimation. In this paper, we formulate a novel block-Kronecker-structured cascaded channel model for a \textit{group-connected} BD-RIS-aided multi-user (MU) mmWave system equipped with uniform planar arrays (UPAs). By exploiting the cascaded channel sparsity, an efficient three-stage estimation protocol is proposed. Specifically, Stage I acquires the common angles of arrival (AoAs) at the base station (BS) via a discrete Fourier transform (DFT)-based approach. Stage II leverages the block-Kronecker structure alongside orthogonal matching pursuit (OMP) and correlation-based least squares (LS) to extract the complete cascaded channel for a designated typical user. Finally, Stage III utilizes a Hierarchical Block OMP (HBOMP) algorithm to estimate the other users' channels. This structurally reconstructs the common and user-specific components, which fundamentally reduces the computational complexity and substantially reduces the pilot overhead. Numerical simulations verify that the proposed protocol yields improved channel estimation accuracy while maintaining a relatively low pilot overhead.
\end{abstract}


\section{Introduction}

\IEEEPARstart{R}{econfigurable} intelligent surface (RIS) is widely recognized as an innovative  paradigm to proactively customize the radio propagation environment for future wireless networks \cite{direnzo2019, pan2021}. 
Comprising an enormous number of low-cost passive reflective elements, 
RIS is capable of artificially altering the phase and amplitude of impinging electromagnetic waves \cite{pan2022overview}. 
This unique capability has spurred its integration into diverse applications, including simultaneous wireless information and power transfer (SWIPT) architectures,
 multicell interference mitigation, and beyond \cite{pan2020mimo}. Particularly in millimeter-wave (mmWave) multiple-input multiple-output
  (MIMO) communications, where signals are vulnerable to severe path attenuation and physical blockages \cite{akdeniz2014},
 integrating an RIS offers a cost-effective remedy to bypass obstacles by constructing programmable virtual line-of-sight (LoS) links.
 Nevertheless, fully unlocking such significant beamforming and spatial multiplexing gains heavily relies on the acquisition of accurate channel state information (CSI) \cite{zhou2022}.
  Given that the passive elements on the RIS are generally lack of active radio frequency (RF) chains and signal processing capability, estimating the high-dimensional cascaded user-RIS-base station (BS) channel imposes a severe pilot overhead.

To alleviate this excessive pilot overhead, extensive research has leveraged the limited scattering nature of high-frequency mmWave channels. Advanced compressive sensing (CS) methodologies and angle-domain signal processing techniques and Bayesian learning frameworks \cite{xie2017, fan2017, zyang2013data} have been successfully extended to RIS-aided systems. By utilizing sparse signal recovery algorithms \cite{tropp2004}, various practical estimators have been developed. For instance, the authors of \cite{wang2020comp} proposed a CS-based method to estimate the cascaded channel by exploiting its inherent sparse structure. For multi-user (MU) scenarios, combining orthogonal matching pursuit (OMP) algorithm, a joint iterative estimator was developed in \cite{chen2023} based on the common row-column-block sparsity shared among all users. Similarly, the double-structured (DS) sparsity of the angular cascaded channels was utilized in \cite{wei2021} to formulate a DS-OMP algorithm. To further achieve a significant reduction in pilot overhead for MU mmWave systems, a correlation-based algorithm was proposed in \cite{zhou2022}. This method explicitly extracts the common angular-domain information shared among users and fully leverages the spatial correlations across different cascaded links to efficiently reconstruct the MU estimation problem. Subsequently, the authors of \cite{peng2022} extended this strategy to uniform planar array (UPA)-type MU-MIMO systems, constructing an efficient three-stage cascaded channel estimation pipeline to systematically resolve the multi-dimensional angles.

Recently, the conventional RIS architecture has been extended to beyond diagonal (BD)-RIS architectures, which inherently adopt non-diagonal phase shift matrices \cite{nerini2024beyond, li2024ris2}. By bridging the passive elements via generalized reconfigurable impedance networks (such as \textit{single-}, \textit{group-}, and \textit{fully-connected} architectures \cite{li2023beyond}), BD-RIS relaxes the diagonal constraint of traditional scattering parameter network models \cite{shen2022}. Unlike conventional RIS where signals are simply reflected by isolated elements, this reciprocal inter-element coupling allows an incident wave on one element to be distributed and re-radiated by other interconnected elements, maximizing the spatial beamforming degrees of freedom (DoF) and extending the localized reflection capability to full-space multi-sector coverage \cite{li2023multisector, zhang2022ios}. Consequently, extensive research has rigorously explored the systemic limits and beamforming capabilities of BD-RIS, yielding advances in discrete-value phase shift designs \cite{nerini2023discrete}, closed-form global optimization bounds \cite{nerini2024closed}, and graph theory-assisted network modeling \cite{nerini2024beyond}.

Among the BD-RIS configurations, the \textit{group-connected} architecture emerges as a highly promising topology for practical deployment. To circumvent \textit{fully-connected} hardware complexities while maintaining substantial spatial beamforming gains, it restricts inter-element connections specifically within disjoint sub-groups. However, this topology fundamentally breaks the canonical diagonal structure of conventional RIS cascaded channels. By generating a multi-port scattering matrix with non-zero off-diagonal entries, the \textit{group-connected} BD-RIS intrinsically mixes incident signals, which tightly couples the angles of arrival (AoAs) and angles of depart (AoDs) at the RIS and renders traditional element-wise estimators ineffective. To address this, recent studies have explored specialized training issues \cite{li2024channel}, advanced tensor decompositions \cite{dealmeida2025channel}, and low-overhead estimation frameworks for MU-MIMO variants \cite{wang2025low}. Nevertheless, these existing methods exhibit distinct limitations. Customized training protocols \cite{li2024channel} primarily rely on unstructured LS estimation of the composite channel. Ignoring the built-in block-Kronecker structure of the cascaded links, they incur a high training overhead that increases heavily with the estimate scale. While tensor decompositions \cite{dealmeida2025channel} successfully exploit multi-dimensional algebraic properties, their inability to structurally isolate common components across users inevitably leads to redundant pilot overhead. 
Furthermore, although the framework in \cite{wang2025low} reduces training overhead, it suffers from error propagation during its sequential estimation process, and additionally, its formulation is strictly confined to the \textit{fully-connected} architecture, lacking generalizability to other topologies. Since
both the \textit{single-connected} and \textit{fully-connected} mathematically
reduce to special cases of the \textit{group-connected} structure, an
estimator tailored for the \textit{group-connected} structure naturally provides general applicability.
Therefore, developing an efficient cascaded channel estimator for \textit{group-connected} BD-RIS that can alleviate the severe pilot overhead in MU settings remains a critical challenge.

In this context, we propose a comprehensive three-stage uplink channel estimation protocol
 tailored for $\textit{group-connected}$ BD-RIS-aided MU mmWave multiple-input single-output (MISO) systems, 
 where both the BS and the RIS are equipped with UPAs. 
 By leveraging the cascaded channel sparsity and exploiting structural characteristics of the common RIS-BS channel shared by all users, 
 this work systematically tackles the non-diagonal structural challenges of BD-RIS. 
 The principal contributions of this paper are outlined as follows:

\begin{itemize}
    \item We propose an efficient three-stage channel estimation protocol for a $\textit{group-connected}$ BD-RIS-aided MU mmWave system based on a multi-port network scattering model. In Stage I, the common AoAs at the BS is proactively acquired using a discrete Fourier transform (DFT)-based method along with an angle rotation technique. In Stage II, By exploiting OMP and leveraging the inherent correlation structure, complete cascaded CSI estimation is performed for the typical user. The common parameters at the RIS are thereby extracted, enabling the reconstructing of channel estimation in an MU scenario. Finally, in Stage III, we utilize these common cascaded parameters with a proposed Hierarchical Block OMP (HBOMP) to drastically reduce the required pilot overhead for estimating the remaining parameters of other users.
    \item We establish a unified block-Kronecker-structured cascaded channel model for $\textit{group-connected}$ BD-RIS under a UPA layout. The model maps the block-diagonal scattering connections of BD-RIS into a structurally sparse array response representation, thereby offering a general and unified framework applicable to a variety of RIS architectures beyond conventional diagonal configurations.
    \item We develop a specialized estimation algorithm combining OMP and a correlation-based least squares (LS) method rigorously tailored for \textit{group-connected} BD-RIS. By exploiting the block-Kronecker structure, the OMP algorithm accurately separates the intertwined AoAs and AoDs at the RIS, and the correlation-based LS retrieves the corresponding channel angles and gains, successfully resolving the full-CSI estimation for the typical user.
    \item A HBOMP approach is executed to successively estimate the cascaded channels of the remaining users. By exploiting the 1-sparse a priori, this approach constrains the search subspace, yielding accurate MU channel estimation with significantly reduced pilot overhead and computational complexity.
\end{itemize}

The subsequent sections of this article are arranged as follows. Section \ref{sec:sysmodel} formulates the comprehensive system model, defining both the subchannels and the architecture of the BD-RIS. Section \ref{sec:channel_estimation} elaborates on the complete design of the proposed cascaded channel estimation protocol. Then, an evaluation regarding the pilot overhead and computational complexity are provided in Section \ref{sec:pilot_and_complexity}. Simulation results validating our theoretical claims are discussed in Section \ref{sec:simulation_results}. The main conclusions of this work are drawn in Section \ref{subsec:conclusion}.

\textit{Notations:} Vectors and matrices are denoted by boldface lowercase and uppercase letters, respectively. For a matrix $\mathbf{X}$ of arbitrary size, the symbols $\mathbf{X}^*$, $\mathbf{X}^{\mathrm{T}}$, and $\mathbf{X}^{\mathrm{H}}$ represent the conjugate, transpose, and Hermitian of matrix $\mathbf{X}$, respectively. $\mathbf{X}^{-1}$ denotes the inverse of a full-rank square matrix $\mathbf{X}$. The Frobenius norm of matrix $\mathbf{X}$ is denoted by $\|\mathbf{X}\|_F$, the Euclidean norm and the $\ell_\infty$-norm of vector $\mathbf{x}$ are denoted by $\|\mathbf{x}\|_2$ and $\|\mathbf{x}\|_{\infty}$, respectively. The modulus of a scalar is denoted by $|\cdot|$. $\Diag(\mathbf{x})$ denotes a diagonal matrix with the entries of vector $\mathbf{x}$ on its main diagonal, and $\blkdiag(\mathbf{X}_1, \ldots, \mathbf{X}_n)$ constructs a block-diagonal matrix with matrices $\mathbf{X}_1, \ldots, \mathbf{X}_n$ on its main diagonal. The vectorization operator $\mathrm{vec}(\mathbf{X})$ stacks the columns of $\mathbf{X}$ into a column vector. The expectation operator is denoted by $\mathbb{E}[\cdot]$. 
$[\mathbf{X}]_{m,n}$ denotes the $(m, n)$-th element of matrix $\mathbf{X}$ and $[\mathbf{x}]_{m}$ denotes the $m$-th element of vector $\mathbf{x}$. The $m$-th row and $n$-th column of matrix $\mathbf{X}$ are represented by $[\mathbf{X}]_{m,:}$ and $[\mathbf{X}]_{:,n}$, respectively. The Kronecker product between two matrices $\mathbf{X}$ and $\mathbf{Y}$ is denoted by $\mathbf{X} \otimes \mathbf{Y}$, and $\lfloor x \rfloor$ rounds down to the nearest integer.

\section{System Model}
\label{sec:sysmodel}
A narrow-band time-division duplex (TDD) mmWave MISO system is considered, where $K$ single-antenna users communicate with a BS equipped with a UPA of $N = N_h \times N_v$ antennas, where $N_h$ and $N_v$ are the number of horizontal and vertical elements, respectively. A BD-RIS, which comprises a passive reflecting UPA of $M = M_h \times M_v$ elements, is deployed to enhance the communication between the users and the BS. The $M$ elements of the BD-RIS are connected to an $M$-port \textit{group-connected} reconfigurable impedance network, 
 where the $M$ ports are uniformly divided into $G$ groups. It is assumed that the direct channels between the BS and the users are blocked, 
 and thus we focus on the estimation of the cascaded user-RIS-BS channels. The following subsections first describe the RIS-BS and user-RIS subchannels, and then present the cascaded channel model while accounting for the BD-RIS architecture.
\subsection{Subchannel Model}
Consider the channel matrix from the RIS to the BS, denoted by $\mathbf{H}\in\mathbb{C}^{N\times M}$, and the channel from the user to the RIS, denoted by $\mathbf{h}_{k}\in\mathbb{C}^{M\times1}$, which are respectively given by

\begin{align}
\mathbf{H} &= \sum_{l=1}^{L}\alpha_{l}\mathbf{a}_{N}\left(\psi_{l},\nu_{l}\right)\mathbf{a}_{M}^{\mathrm{H}}\left(\omega_{l},\mu_{l}\right),
\label{eq:H_RIS_BS}\\
\mathbf{h}_{k} &= \sum_{j=1}^{J_{k}}\beta_{k,j}\mathbf{a}_{M}\left(\varphi_{k,j},\theta_{k,j}\right),\quad \forall k \in \mathcal{K},
\label{eq:h_user_RIS}
\end{align}
where $L$ and $J_k$ denote the number of propagation paths between the BS and the RIS, and 
between the RIS and user $k$, respectively. For the $l$-th RIS-BS path, 
$\alpha_{l}$ is the complex gain, $(\psi_{l}, \nu_{l})$ and $(\omega_{l}, \mu_{l})$ are the spatial frequencies corresponding 
to the azimuth and elevation AoAs at the BS, and AoDs at the RIS, respectively. Similarly, for the $j$-th user-RIS path, 
 $\beta_{k,j}$ is the complex gain, and $(\varphi_{k,j},\theta_{k,j})$ are the spatial frequencies for the AoAs 
 at the RIS from user $k$. The set of users is defined as $\mathcal{K} = \{1,2,\ldots, K\}$.

The Saleh--Valenzuela (SV) model is adopted \cite{peng2022} to characterize the channel to exploit the limited scattering nature of mmWave propagation, and the array response vector (ARV) for a UPA of size $P = P_h \times P_v$ can be represented by
\begin{align}
\mathbf{a}_{P}(z,x) = \mathbf{a}_{P_v}(z) \otimes \mathbf{a}_{P_h}(x),
\label{eq:UPA_response}
\end{align}
where $\mathbf{a}_{P_v}(z)$ and $\mathbf{a}_{P_h}(x)$ denote the steering vectors with respect to the vertical ($z$-axis) and horizontal ($x$-axis) of the UPA, which are respectively given by
\begin{align}
\mathbf{a}_{P_v}(z) &= [1, e^{-\mathrm{j}2\pi z}, \ldots, e^{-\mathrm{j}2\pi(P_v-1)z}]^{\mathrm{T}}, \nonumber\\
\mathbf{a}_{P_h}(x) &= [1, e^{-\mathrm{j}2\pi x}, \ldots, e^{-\mathrm{j}2\pi(P_h-1)x}]^{\mathrm{T}}. 
\end{align}
The spatial frequency pair $(z,x)$ is defined based on
 the physical elevation angle $\rho \in [-90^\circ,90^\circ)$ and the azimuth angle $\xi \in 
 [-180^\circ,180^\circ)$ with the relationship as follows, $z = \frac{d}{\lambda_c}\cos(\rho)$ and $x = \frac{d}{\lambda_c}\sin(\rho)\cos(\xi)$, 
where $\lambda_c$ is the carrier wavelength and $d$ is the inter-element spacing of the UPA.
The channel in Eq.~\eqref{eq:H_RIS_BS} can be compactly written in matrix form as
\begin{align}
\mathbf{H} = \mathbf{A}_{N}\boldsymbol{\Lambda}\mathbf{A}_{M}^{\mathrm{H}},
\label{eq:H_matrix_form}
\end{align}
where
\begin{subequations}
\begin{align}
\mathbf{A}_{N} &= [\mathbf{a}_{N}(\psi_{1},\nu_{1}), \ldots, \mathbf{a}_{N}(\psi_{L},\nu_{L})] \in \mathbb{C}^{N \times L}, \label{eq:A_N_def}\\
\boldsymbol{\Lambda} &= \mathrm{Diag}(\alpha_{1}, \ldots, \alpha_{L}) \in \mathbb{C}^{L \times L}, \label{eq:Lambda_def}\\
\mathbf{A}_{M} &= [\mathbf{a}_{M}(\omega_{1},\mu_{1}), \ldots, \mathbf{a}_{M}(\omega_{L},\mu_{L})] \in \mathbb{C}^{M \times L}. \label{eq:A_M_def}
\end{align}
\end{subequations}
The $k$-th user-RIS channel in Eq.~\eqref{eq:h_user_RIS} can also be expressed in a more compact form as
\begin{align}
\mathbf{h}_{k} = \mathbf{A}_{M,k}\boldsymbol{\beta}_{k}, \quad \forall k \in \mathcal{K},
\label{eq:hk_matrix_form}
\end{align}
where
\begin{subequations}
\begin{align}
\mathbf{A}_{M,k} &= [\mathbf{a}_{M}(\varphi_{k,1},\theta_{k,1}), \ldots, \mathbf{a}_{M}(\varphi_{k,J_{k}},\theta_{k,J_{k}})] \in \mathbb{C}^{M \times J_{k}}, \label{eq:A_Mk_def}\\
\boldsymbol{\beta}_{k} &= [\beta_{k,1}, \ldots, \beta_{k,J_{k}}]^{\mathrm{T}} \in \mathbb{C}^{J_{k} \times 1}. \label{eq:beta_k_def}
\end{align}
\end{subequations}
\subsection{BD-RIS Scattering Matrix Model}
Consider a BD-RIS operating in reflective mode. It is distinguished from the conventional diagonal RIS by the inter-element connections, which 
render its scattering matrix non-diagonal. Our study centers on the \textit{group-connected} architecture, which is characterized by partitioning the $M$ RIS elements uniformly into $G$ groups, with $\bar{M} = M/G$ interconnected elements per group. Accordingly, the scattering matrix $\mathbf{\Phi} \in \mathbb{C}^{M \times M}$ possesses a block-diagonal structure, given by
\begin{align}
\mathbf{\Phi}=\mathrm{blkdiag}(\overline{\mathbf{\Phi}}_{1},\ldots,\overline{\mathbf{\Phi}}_{G}),
\label{eq:Phi}
\end{align}
where each block $\overline{\mathbf{\Phi}}_{g} \in \mathbb{C}^{\bar{M} \times \bar{M}}, \forall g \in \{1,\ldots, G\}$ is a unitary matrix, i.e., $\overline{\mathbf{\Phi}}_{g}^{\mathrm{H}} \overline{\mathbf{\Phi}}_{g} = \mathbf{I}_{\bar{M}}$. The scattering matrix $\mathbf{\Phi}$ can also be determined by its admittance matrix $\mathbf{Y} \in \mathbb{C}^{M \times M}$ as $\mathbf{\Phi}=(\mathbf{I}+Z_0\mathbf{Y})^{-1}(\mathbf{I}-Z_0\mathbf{Y})$, where $Z_0$ is the characteristic impedance and $\mathbf{Y}$ also exhibits a block-diagonal structure.

The \textit{single-connected} (diagonal) and \textit{fully-connected} architectures can be viewed as special cases of the \textit{group-connected} structure by adjusting the number of groups $G$. For the \textit{fully-connected} case, $G=1$, leading to $\mathbf{\Phi}=\overline{\mathbf{\Phi}}_{1}$, where $\overline{\mathbf{\Phi}}_{1}$ is a full unitary matrix. For the \textit{single-connected} case, $G=M$, resulting in $\mathbf{\Phi}=\mathrm{Diag}(e_1,\ldots,e_M)$ with $|e_g|=1, \forall g$. Other connection patterns, such as \textit{tree-connected} or \textit{forest-connected}, result in a scattering matrix $\mathbf{\Phi}$ whose non-zero element positions are not predetermined and can only be characterized through the properties of its admittance matrix $\mathbf{Y}$; in such cases, $\mathbf{\Phi}$ can be considered an arbitrary matrix.
In the remainder of this paper, the BD-RIS with a scattering matrix in the form of Eq.~\eqref{eq:Phi} is adopted.
\subsection{Cascaded Channel Model}
The traditional cascaded channel model for a conventional RIS is given by $\mathbf{h}_{\mathrm{cv},k}=\mathbf{H}\mathbf{\Phi}\mathbf{h}_{k}$, where the scattering matrix $\mathbf{\Phi}$ corresponds to the \textit{single-connected} structure mentioned above. However, this model is inapplicable to the BD-RIS structure and the traditional cascaded channel formulation cannot be directly applied. As for BD-RIS, the scattering matrix $\mathbf{\Phi}$ is considered in Eq.~\eqref{eq:Phi}.

To extend this model to the \textit{group-connected} BD-RIS, the columns of steering matrices at the RIS in Eq.~\eqref{eq:A_M_def} and Eq.~\eqref{eq:A_Mk_def} are rearranged to align with the group structure. Consequently, the matrices can be rewritten as $\mathbf{A}_{M} = [(\mathbf{a}_{M})_{\mathrm{ra}}(\omega_{1},\mu_{1}), \ldots, (\mathbf{a}_{M})_{\mathrm{ra}}(\omega_{L},\mu_{L})] \in \mathbb{C}^{M \times L}, \mathbf{A}_{M,k} = [(\mathbf{a}_{M})_{\mathrm{ra}}(\varphi_{k,1},\theta_{k,1}), \ldots, (\mathbf{a}_{M})_{\mathrm{ra}}(\varphi_{k,J_{k}},\theta_{k,J_{k}})] \in \mathbb{C}^{M \times J_{k}}$.
Since the form of the both matrices is irrelevant with the work, we assume it is arranged in a common way that is grouped into identical squares, and both the groups and the elements within each group are numbered sequentially, following a two-level horizontal-then-vertical order. Then the structure of $(\mathbf{a}_{M})_{\mathrm{ra}}(z,x)$ is elucidated, as illustrated below:
\begin{align}
[(\mathbf{a}_{M})_{\mathrm{ra}}(z,x)]_{i} = [\mathbf{a}_{M}(z,x)]_{j},
\end{align}
where $
j = \bar{M} M_v \left\lfloor \frac{i}{\bar{M} M_v} \right\rfloor
    + M_v \left( \left\lfloor \frac{i}{\bar{M}} \right\rfloor \bmod \bar{M} \right)
    + (i \bmod \bar{M})
    + \bar{M} \left( \left\lfloor \frac{i}{\bar{M}^2} \right\rfloor \bmod \frac{M_v}{\bar{M}} \right).
$

Therefore, in the remainder of this paper, we uniformly assume that $(\mathbf{a}_{M})(z,x)$ refers to $(\mathbf{a}_{M})_{\mathrm{ra}}(z,x)$ of this form.
To simplify the cascaded channel model, we exploit its block-diagonal structure to rewrite the high-dimensional matrix product as a sum of lower-dimensional products. 
Specifically,
\begin{align}
&\mathbf{h}_{\mathrm{BD},k} = \mathbf{H}\mathbf{\Phi}\mathbf{h}_{k} = \mathbf{H} \times \mathrm{blkdiag}(\overline{\mathbf{\Phi}}_{1},\ldots,\overline{\mathbf{\Phi}}_{G}) \times \mathbf{h}_{k} \nonumber\\
&= \begin{bmatrix}
\mathbf{H}_1,\mathbf{H}_2,\ldots,\mathbf{H}_G
\end{bmatrix}
\begin{bmatrix}
\overline{\mathbf{\Phi}}_{1} & & & \\
& \overline{\mathbf{\Phi}}_{2} & & \\
& & \ddots & \\
& & & \overline{\mathbf{\Phi}}_{G}
\end{bmatrix}
\begin{bmatrix}
\mathbf{h}_{k,1}\\
\mathbf{h}_{k,2}\\
\vdots \\
\mathbf{h}_{k,G}
\end{bmatrix} \nonumber\\&= \sum_{g=1}^{G}\mathbf{H}_g\overline{\mathbf{\Phi}}_{g}\mathbf{h}_{k,g},\label{eq:cascaded_group_sum}
\end{align}
where $\mathbf{H}_g=[\mathbf{H}]_{1+(g-1)\bar{M}:g\bar{M},:}$ and $\mathbf{h}_{k,g}=[\mathbf{h}_{k}]_{1+(g-1)\bar{M}:g\bar{M}}$ are partitioned according to the RIS grouping. Unless otherwise specified, the subscript $g$ in the following refers to blocks grouped in the same manner.

Consequently, Eq.~\eqref{eq:cascaded_group_sum} can be equivalently reformulated into the following compact form:
\begin{align}
&\mathbf{h}_{\mathrm{BD},k} = \mathrm{vec}(\mathbf{H}\mathbf{\Phi}\mathbf{h}_{k}) = \mathrm{vec}\left( \sum_{g=1}^{G}\mathbf{H}_g\overline{\mathbf{\Phi}}_{g}\mathbf{h}_{k,g} \right) \nonumber \\
&= \sum_{g=1}^{G} \mathrm{vec}(\mathbf{H}_g\overline{\mathbf{\Phi}}_{g}\mathbf{h}_{k,g}) = \sum_{g=1}^{G} (\mathbf{h}_{k,g}^{\mathrm{T}} \otimes \mathbf{H}_g) \mathrm{vec}(\overline{\mathbf{\Phi}}_{g}) \nonumber \\
&= \underbrace{[\mathbf{h}_{k,1}^{\mathrm{T}} \otimes \mathbf{H}_1, \ldots, \mathbf{h}_{k,G}^{\mathrm{T}} \otimes \mathbf{H}_G]}_{\bar{\mathbf{B}}_k \in \mathbb{C}^{N \times \bar{M}^2 G}} \underbrace{
\begin{bmatrix}
\mathrm{vec}(\overline{\mathbf{\Phi}}_1) \\
\mathrm{vec}(\overline{\mathbf{\Phi}}_2) \\
\vdots \\
\mathrm{vec}(\overline{\mathbf{\Phi}}_{G})
\end{bmatrix}
}_{\bar{\mathbf{p}} \in \mathbb{C}^{\bar{M}^2 G \times 1}} = \bar{\mathbf{B}}_k\bar{\mathbf{p}},
\label{eq:cascaded_vec_form}
\end{align}
where $\bar{\mathbf{p}}$ is the vectorized phase shift matrix for BD-RIS. By substituting the geometric models $\mathbf{H}_g = \mathbf{A}_N \boldsymbol{\Lambda} (\mathbf{A}_M^{\mathrm{H}})_g$ and $\mathbf{h}_{k,g} = (\mathbf{A}_{M,k})_g \boldsymbol{\beta}_k$ into $\bar{\mathbf{B}}_{k,g}\triangleq\mathbf{h}_{k,g}^{\mathrm{T}} \otimes \mathbf{H}_g$, and exploiting the mixed-product property of the Kronecker product, i.e., $(\mathbf{AB})\otimes(\mathbf{CD})=(\mathbf{A}\otimes \mathbf{C})(\mathbf{B}\otimes \mathbf{D})$, we have
\begin{align}
\mathbf{\bar{B}}_{k,g} &= \mathbf{h}_{k,g}^{\mathrm{T}} \otimes [\mathbf{A}_N \boldsymbol{\Lambda} (\mathbf{A}_M^{\mathrm{H}})_g] \nonumber \\
&= (\boldsymbol{\beta}_k^{\mathrm{T}}(\mathbf{A}_{M,k})_g^{\mathrm{T}}) \otimes [\mathbf{A}_N \boldsymbol{\Lambda} (\mathbf{A}_M^{\mathrm{H}})_g] \nonumber \\
&= (\boldsymbol{\beta}_k^{\mathrm{T}} \otimes (\mathbf{A}_N\boldsymbol{\Lambda}))((\mathbf{A}_{M,k})_g^{\mathrm{T}} \otimes (\mathbf{A}_M^{\mathrm{H}})_g)\nonumber\\
&= \mathbf{A}_N (\boldsymbol{\beta}_k^{\mathrm{T}} \otimes \boldsymbol{\Lambda}) ((\mathbf{A}_{M,k})_g^{\mathrm{T}} \otimes (\mathbf{A}_M^{\mathrm{H}})_g).
\end{align}
Therefore, the overall cascaded channel can be expressed as
\begin{align}
&\mathbf{h}_{\mathrm{BD},k} = \mathbf{A}_N (\boldsymbol{\beta}_k^{\mathrm{T}} \otimes \boldsymbol{\Lambda})\nonumber\\
&\underbrace{[(\mathbf{A}_{M,k})_1^{\mathrm{T}} \otimes (\mathbf{A}_M^{\mathrm{H}})_1, \ldots, (\mathbf{A}_{M,k})_G^{\mathrm{T}} \otimes (\mathbf{A}_M^{\mathrm{H}})_G]}_{\mathbf{B}_k \in \mathbb{C}^{J_{k}L \times \bar{M}^2 G}}
\bar{\mathbf{p}},
\label{eq:cascaded_final_form}
\end{align}
where $\mathbf{B}_k$ denotes the composite block-Kronecker-structured RIS steering matrix. The distinctive grouped structure of this matrix renders the estimation problem fundamentally different from that of the conventional cascaded channel, which constitutes the primary focus of this work.

\subsection{Received Signal}
Let $s_k(t)$ denote the pilot signal transmitted by user $k$. The received signal at the BS for user $k$ at time slot $t$ ($1 \le t \le \tau_k$) is given by
\begin{align}
\mathbf{y}_{k}(t) = \sqrt{p} \, \mathbf{A}_N (\boldsymbol{\beta}_k^{\mathrm{T}} \otimes \boldsymbol{\Lambda}) \mathbf{B}_k \bar{\mathbf{p}}_t \, s_{k}(t) + \mathbf{n}_{k}(t), \quad \forall k \in \mathcal{K},
\label{eq:received_signal_slot}
\end{align}
where $\bar{\mathbf{p}}_{t}$ is the RIS equivalent phase shift vector at time slot $t$, and $\mathbf{n}_{k}(t) \in \mathbb{C}^{N\times1} \sim \mathcal{CN}(\mathbf{0}, \delta^{2}\mathbf{I})$ is the additive white Gaussian noise (AWGN). Assuming the pilot symbols satisfy $s_{k}(t)=1$ for $1 \leq t \leq \tau_{k}$, we stack the received signals over $\tau_k$ time slots to form the measurement matrix
\begin{align}
\mathbf{Y}_{k} = [\mathbf{y}_{k}(1), \ldots, \mathbf{y}_{k}(\tau_k)]  = \sqrt{p} \, \mathbf{G}_{k} \boldsymbol{\Theta}_{k} + \mathbf{N}_{k},
\label{eq:stacked_measurement}
\end{align}
where $\mathbf{Y}_{k} \in \mathbb{C}^{N \times \tau_k}$ denotes the received signal matrix, $\boldsymbol{\Theta}_{k} = [\bar{\mathbf{p}}_1, \ldots, \bar{\mathbf{p}}_{\tau_k}] \in \mathbb{C}^{\bar{M}^2 G \times \tau_k}$ denotes the equivalent RIS phase shift matrix, and $\mathbf{N}_{k} = [\mathbf{n}_{k}(1), \ldots, \mathbf{n}_{k}(\tau_k)]$ denotes the noise matrix, and
$\mathbf{G}_{k} \triangleq \mathbf{A}_N (\boldsymbol{\beta}_k^{\mathrm{T}} \otimes \boldsymbol{\Lambda}) \mathbf{B}_k \in \mathbb{C}^{N \times \bar{M}^2 G}$
is defined as the equivalent cascaded channel, which needs to be estimated. 
While this model explicitly represents the \textit{group-connected} property of the BD-RIS by splitting subchannels into groups, the resulting group structure of $\mathbf{B}_k$ inherently leads to a loss of spatial correlation among groups. This structural limitation poses significant challenges for channel estimation, particularly in the MU phase, as will be discussed in Section~\ref{subsec:est_other_users}.
\begin{figure*}[htbp]
    \centering
    \includegraphics[width=\textwidth]{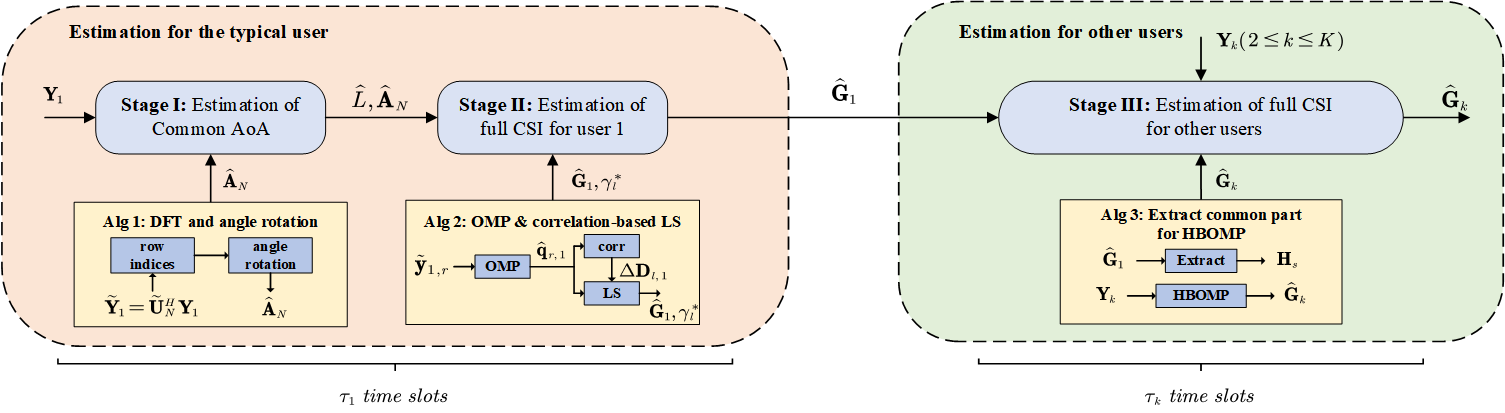}
    \caption{The proposed three-stage channel estimation protocol for BD-RIS.}
    \label{fig:1}
\end{figure*}
\section{Channel Estimation}
\label{sec:channel_estimation}
\subsection{Channel Estimation Protocol}

The proposed three-stage uplink channel estimation protocol for BD-RIS is illustrated in Fig.~\ref{fig:1}. Existing \textit{group-connected} BD-RIS channel estimation methods \cite{li2024channel,dealmeida2025channel} remain limited in MU systems because the inter-element coupling tightly entangles the involved spatial angles, making it difficult to isolate common structure. To address this issue, our protocol follows a staged estimation framework that identifies common components, refines user-specific cascaded-channel components, and reuses the extracted priors for structured recovery of the other users over a reduced search space, thereby reducing the pilot overhead and computational complexity.

\textit{Stage I: Common AoA Estimation at the BS.} In this stage, we estimate the common AoA matrix at the BS, which is shared by all users due to the common BS-RIS channel. We use the DFT-based method to extract the dominant spatial frequencies. To mitigate the power leakage effect caused by finite values of $N_v$ and $N_h$, we introduce a low-complexity angle rotation technique, which improves the estimation accuracy with minimal additional computational overhead.

\textit{Stage II: Cascaded CSI Estimation for the Typical User.} After obtaining the common AoA, we focus on recovering the BD-RIS cascaded channel matrix for the typical user, denoted as user~1. We select the user closest to the RIS as the typical user, which provides a high Signal-to-Noise Ratio (SNR) for more reliable estimation. However, directly estimating this high-dimensional cascaded channel is computationally expensive. Therefore, we use the block-Kronecker product structure to simplify the process. We first apply the OMP algorithm to extract the maximum-power cascaded path, which serves as a reference column. Then, by exploiting the linear transformation relationship between this reference and the remaining paths, we obtain the complete angular information through correlation-based LS estimation.

\textit{Stage III: CSI Estimation for the Other Users.} In the final stage, we estimate the channels of the remaining $K-1$ users. Rather than estimating each user's CSI independently, we reconstruct their cascaded channels into a common part, shared with the typical user, and user-specific parts. With the reconstruction, sparse recovery over the composite dictionary can still be challenging due to mutual coherence. To address this, we propose the HBOMP algorithm. HBOMP utilizes the intra-block sparsity caused by the 1-sparse geometric alignment of the common channel to restrict the search dimension, which improves the efficiency of the CSI recovery for each user. The details are provided in the following subsections.
\subsection{Stage I:Estimation common AOA at the BS}
\label{subsec:bs_aoa_estimation}
The received signal matrix in Eq.~$\eqref{eq:stacked_measurement}$, $\mathbf{Y}_k \in \mathbb{C}^{N \times \tau_k}$, can be expressed as
\begin{align}
    \mathbf{Y}_k = \sqrt{p}\mathbf{G}_k\mathbf{\Theta}_k+\mathbf{N}_k \triangleq \sqrt{p}\mathbf{A}_N\mathbf{P}_k+\mathbf{N}_k,
\end{align}
where ${\mathbf{P}}_k = (\boldsymbol{\beta}_k^{\mathrm{T}} \otimes \boldsymbol{\Lambda}) [(\mathbf{A}_{M,k})_1^{\mathrm{T}} \otimes (\mathbf{A}_M^{\mathrm{H}})_1, \ldots, (\mathbf{A}_{M,k})_G^{\mathrm{T}} \otimes (\mathbf{A}_M^{\mathrm{H}})_G]$ represents the equivalent received signal matrix for user $k$. Leveraging the characteristics of large-scale antenna arrays and the cascaded channel structure, the DFT-based estimation is conducted as follows.
\subsubsection{Estimation for Common AoAs at the BS}
Define an equivalent Fourier matrix $\widetilde{\mathbf{U}}_N \triangleq \mathbf{U}_{N_h} \otimes \mathbf{U}_{N_v} \in \mathbb{C}^{N \times N}$, where $\mathbf{U}_{N_h}$ and $\mathbf{U}_{N_v}$ are standard DFT matrices. Based on the asymptotic orthogonality of large-scale antenna arrays \cite{peng2022}, the linear transformation $\widetilde{\mathbf{U}}_N^{\mathrm{H}} \mathbf{A}_N$ becomes a tall, sparse matrix. Specifically, each of its $L$ columns contains only one nonzero element. The index $n_l$ of this nonzero element in the $l$-th column is uniquely mapped to the AoA spatial frequency pair $(\psi_l, \nu_l)$ of the $l$-th path as follows:
\begin{subequations}
\begin{align}
    n_h(l) &= \begin{cases}
        N_h \psi_l + 1, & \psi_l \in [0, \frac{d_{\mathrm{BS}}}{\lambda_c}) \\
        N_h + N_h \psi_l + 1, & \psi_l \in [-\frac{d_{\mathrm{BS}}}{\lambda_c}, 0)
    \end{cases}, \\
    n_v(l) &= \begin{cases}
        N_v \nu_l + 1, & \nu_l \in [0, \frac{d_{\mathrm{BS}}}{\lambda_c}) \\
        N_v + N_v \nu_l + 1, & \nu_l \in [-\frac{d_{\mathrm{BS}}}{\lambda_c}, 0)
    \end{cases}, \\
    n_l &= (n_h(l)-1)N_v + n_v(l),
\end{align}
\end{subequations}
where $n_h(l)$ and $n_v(l)$ are integers. The common AoAs can therefore be estimated by identifying the rows of $\widetilde{\mathbf{Y}}_1 = \widetilde{\mathbf{U}}_N^{\mathrm{H}} \mathbf{Y}_1$ with significant power. Let $\Omega_N = \{n_l\}_{l=1}^{\widehat{L}}$ be the set of indices for these $\widehat{L}$ rows. The corresponding antenna indices $(n_h(l), n_v(l))$ can be recovered from $n_l$ using:
\begin{equation}
    n_h(l) = \left\lceil \frac{n_l}{N_v} \right\rceil, \quad n_v(l) = n_l - N_v(n_h(l)-1).
    \label{eq:index_mapping}
\end{equation}
The initial coarse estimates of the AoA spatial frequencies are given by $\widehat{\psi}_l^{init} = (n_h(l)-1)/N_h$ and $\widehat{\nu}_l^{init} = (n_v(l)-1)/N_v$.

\subsubsection{Low-Complexity Angle Rotation for Suppressing Power Leakage}
\label{subsubsec:angle_rotation}

In practice, due to the finite number of antennas, the DFT-based estimation suffers from \textit{power leakage} \cite{wang2019squint}, which degrades the estimation accuracy. To concentrate the leaked power and refine the angle estimates, an angle rotation operation is performed for each estimated AoA pair.

For the $l$-th AoA pair, a rotation matrix is defined as $\mathbf{R}(\Delta\psi_l, \Delta\nu_l) = \mathbf{R}_1(\Delta\psi_l) \otimes \mathbf{R}_2(\Delta\nu_l)$, where $\mathbf{R}_1$ and $\mathbf{R}_2$ are diagonal phase rotation matrices, i.e.
\begin{align}
    \mathbf{R}_1(\Delta\psi_l)&=\operatorname{Diag}\{1,e^{-\mathrm{j}\Delta\psi_l},\ldots,e^{-\mathrm{j}(N_h-1)\Delta\psi_l}\},\\
    \mathbf{R}_2(\Delta\nu_l)&=\operatorname{Diag}\{1,e^{-\mathrm{j}\Delta\nu_l},\ldots,e^{-\mathrm{j}(N_v-1)\Delta\nu_l}\}.
\end{align}
The optimal rotation parameters $(\Delta\widehat{\psi}_l, \Delta\widehat{\nu}_l)$ that compensate for the leakage can be found via a two-dimensional (2-D) search over a small region:
\begin{align}
(\Delta\widehat{\psi}_{l},\Delta\widehat{\nu}_{l})=\arg\max_{\Delta\psi_{l}\in[-\frac{\pi}{N_{h}},\frac{\pi}{N_{h}}], \Delta\nu_{l}\in[-\frac{\pi}{N_{v}},\frac{\pi}{N_{v}}]} \nonumber\\ \left\|[\widetilde{\mathbf{U}}_{N}]^{\mathrm{H}}_{:,n_{l}}\mathbf{R}(\Delta\psi_{l},\Delta\nu_{l})\mathbf{Y}_{1}\right\|^{2}.
\label{eq:1D_search}
\end{align}
A 2-D search has a computational complexity of $\mathcal{O}(L g_1 g_2)$, where $g_1$ and $g_2$ denote the grid sizes. Drawing the decomposed procedure established in \cite{peng2022}, we can significantly reduce this overhead by decomposing the 2-D search into two independent 1-D directional searches. This decomposed algorithm is realized by constructing two specialized rotation matrices, defined as $\widetilde{\mathbf{R}}_1(\Delta\psi) \triangleq \mathbf{R}_1(\Delta\psi) \otimes \mathbf{J}_{N_v}$ and $\widetilde{\mathbf{R}}_2(\Delta\nu) \triangleq \mathbf{J}_{N_h} \otimes \mathbf{R}_2(\Delta\nu)$. Here, the auxiliary selection matrix $\mathbf{J}_N \triangleq \Diag(1, 0, \ldots, 0) \in \mathbb{R}^{N \times N}$ is a square diagonal matrix. It serves to extract the leading component along the specified dimension. With this construction, the optimal 1-D rotation parameters can be efficiently determined through two separate 1-D searches:
\begin{align}
\Delta\widehat{\psi}_l &= \arg\max_{\Delta\psi_l\in\left[-\frac{\pi}{N_h},\frac{\pi}{N_h}\right]}\left\|\left[\widetilde{\mathbf{U}}_1\right]_{:,\overline{n_{1 l}}}^{\mathrm{H}}\widetilde{\mathbf{R}}_1\left(\Delta\psi_l\right) \mathbf{Y}_1\right\|^2, \\
\Delta\widehat{\nu}_l &= \arg\max_{\Delta\nu_l\in\left[-\frac{\pi}{N_v},\frac{\pi}{N_v}\right]}\left\|\left[\widetilde{\mathbf{U}}_2\right]_{:,\overline{n_{2 l}}}^{\mathrm{H}}\widetilde{\mathbf{R}}_2\left(\Delta\nu_l\right) \mathbf{Y}_1\right\|^2,
\end{align}
where the re-indexed coordinate offsets are computed as $\overline{n_{1 l}} = (n_h(l)-1) N_v + 1$ and $\overline{n_{2 l}} = n_v(l)$, respectively. The decomposed transformation matrices $\widetilde{\mathbf{U}}_1$ and $\widetilde{\mathbf{U}}_2$ are defined interchangeably based on the corresponding dimension of the 2-D equivalent Fourier matrix. By substituting this independent angle search procedure, the prohibitive 2-D peak-finding step complexity drops dramatically from $\mathcal{O}(L g_1 g_2)$ down to $\mathcal{O}\left(L\left(g_{1}+g_{2}\right)\right)$. Thus, leveraging these individual rotation shifts $\Delta\widehat{\psi}_l$ and $\Delta\widehat{\nu}_l$, the final refined estimation values for the AoA spatial frequencies are computed by:
\begin{align}
\widehat{\psi}_l &= \begin{cases}\frac{n_h(l)-1}{N_h}-\frac{\Delta\widehat{\psi}_l}{2\pi}& n_h(l)\leq N_h\frac{d_{\mathrm{BS}}}{\lambda_c}\\
\frac{n_h(l)-1}{N_h}-1-\frac{\Delta\widehat{\psi}_l}{2\pi}& n_h(l)>N_h\frac{d_{\mathrm{BS}}}{\lambda_c},\end{cases} \label{eq:psi_estimation}\\
\widehat{\nu}_l &=\left\{\begin{array}{ll}\frac{n_v(l)-1}{N_v}-\frac{\Delta\widehat{\nu}_l}{2\pi}& n_v(l)\leq N_v\frac{d_{\mathrm{BS}}}{\lambda_c}\\
\frac{n_v(l)-1}{N_v}-1-\frac{\Delta\widehat{\nu}_l}{2\pi}& n_v(l)>N_v\frac{d_{\mathrm{BS}}}{\lambda_c}.\end{array}\right.
\label{eq:nu_estimation}
\end{align}
The estimated common AoA steering matrix is $\widehat{\mathbf{A}}_{N}=[\mathbf{a}_{N}(\widehat{\psi}_{1},\widehat{\nu}_{1}),\ldots, \mathbf{a}_{N}(\widehat{\psi}_{\widehat{L}},\widehat{\nu}_{\widehat{L}})]$. The overall procedure is summarized in Algorithm \ref{alg:aoa_estimation}.

\begin{algorithm}
\caption{Low-Complexity Angle Rotation Based AoA Estimation}
\label{alg:aoa_estimation}
\begin{algorithmic}[1]
\Require Received signal matrix $\mathbf{Y}_{1}$.
\State Compute linear transformation: $\widetilde{\mathbf{Y}}_1=\widetilde{\mathbf{U}}_N^{\mathrm{H}} \mathbf{Y}_1$.
\State Calculate row power: $z(n)=\left\|\left[\widetilde{\mathbf{Y}}_1\right]_{n,:}\right\|_2,\forall n$.
\State Find peak power rows: $\left(\Omega_N,\widehat{L}\right)$, where $\Omega_{N}=\{n_{l}\}_{l=1}^{\widehat{L}}$.
\State Construct index sets $\Omega_{N_h}=\{n_h(l)\}, \Omega_{N_v}= \{n_v(l)\}$ via Eq.~(\ref{eq:index_mapping}).
\For{$l=1$ to $\widehat{L}$}
    \State Calculate $\overline{n_{1 l}}$ and $\overline{n_{2 l}}$ via Eq.~$\eqref{eq:index_mapping}$.
    \State Find $\Delta\widehat{\psi}_{l}$ and $\Delta\widehat{\nu}_{l}$ via 1-D searches via Eq.~$\eqref{eq:1D_search}$.
    \State Estimate $\widehat{\psi}_{l}$ and $\widehat{\nu}_{l}$ via Eq.~$\eqref{eq:psi_estimation}$ and Eq.~$\eqref{eq:nu_estimation}$.
\EndFor
\Ensure $\{(\widehat{\psi}_{l},\widehat{\nu}_{l})\}_{l=1}^{\widehat{L}}$ and $\widehat{\mathbf{A}}_{N}$.
\end{algorithmic}
\end{algorithm}

\subsection{Stage II: Estimation of Full CSI for Typical User}
\label{subsec:cascaded_est_path1}
$\textit{1) Estimation of the reference column:} $
Define $\Delta \mathbf{A}_N \triangleq \widehat{\mathbf{A}}_N - \mathbf{A}_N$ as the estimation error of the common AoA. By exploiting the spatial projection over $\mathbf{Y}_k$ and leveraging the asymptotic orthogonality of the massive steering matrices to eliminate the influence of $\mathbf{A}_N$, we have $\widehat{\mathbf{A}}_{N}^{\mathrm{H}} \mathbf{A}_{N}=N \mathbf{I}_{L}+\left(\Delta \mathbf{A}_{N}\right)^{\mathrm{H}} \mathbf{A}_{N}$, and thus we can obtain
\begin{align}
    \frac{1}{N\sqrt{p}}\mathbf{A}_N^{\mathrm{H}} \mathbf{Y}_{k}=(\boldsymbol{\beta}_k^{\mathrm{T}} \otimes \boldsymbol{\Lambda}) \mathbf{B}_k+\tilde{\mathbf{N}}_k^H,
\label{eq:eliminate}
\end{align}
where $\tilde{\mathbf{N}}_k \triangleq \frac{1}{N\sqrt{p}}\mathbf{A}_N^{\mathrm{H}} \mathbf{N}_k + \frac{1}{N}(\Delta\mathbf{A}_N)^{\mathrm{H}}\mathbf{A}_N(\boldsymbol{\beta}_k^{\mathrm{T}} \otimes \boldsymbol{\Lambda})\mathbf{B}_k\mathbf{\Theta}_k$ denotes the corresponding noise, and the second term characterizes the residual error propagation introduced by the common AoA estimation. Let the Hermitian of $\frac{1}{N\sqrt{p}}\mathbf{A}_N^{\mathrm{H}} \mathbf{Y}_{k}$ represent the equivalent measurement matrix, i.e.
\begin{align}
    \tilde{\mathbf{Y}}_{k} & = \left(\frac{1}{N\sqrt{p}}\mathbf{A}_N^{\mathrm{H}} \mathbf{Y}_{k}\right)^{\mathrm{H}} = \mathbf{\Theta}_{k}^{\mathrm{H}} \mathbf{B}_k^{\mathrm{H}} (\boldsymbol{\beta}_k^* \otimes \boldsymbol{\Lambda}^{\mathrm{H}}) + \tilde{\mathbf{N}}_{k} \nonumber \\
    & =\mathbf{\Theta}_{k}^{\mathrm{H}}\left[ \begin{array}{@{}c@{}}
    (\mathbf{A}_{M,k})_1^* \otimes (\mathbf{A}_M)_1 \\
    (\mathbf{A}_{M,k})_2^* \otimes (\mathbf{A}_M)_2 \\
    \vdots \\
    (\mathbf{A}_{M,k})_G^* \otimes (\mathbf{A}_M)_G
    \end{array} \right] (\boldsymbol{\beta}_k^* \otimes \boldsymbol{\Lambda}^{\mathrm{H}})+\tilde{\mathbf{N}}_{k}.
\label{eq:esttrans}
\end{align}
By extracting the $r$-th column of the equivalent measurement matrix $\tilde{\mathbf{Y}}_{1}$, which corresponds to the $r$-th path among the $L$ cascaded paths, we obtain:
\begin{align}
    \tilde{\mathbf{y}}_{1,r} &=\mathbf{\Theta}_{k}^{\mathrm{H}}\left[ \begin{array}{@{}c@{}}
    (\mathbf{A}_{M,1})_1^* \otimes (\mathbf{A}_M)_1 \\
    (\mathbf{A}_{M,1})_2^* \otimes (\mathbf{A}_M)_2 \\
    \vdots \\
    (\mathbf{A}_{M,1})_G^* \otimes (\mathbf{A}_M)_G
    \end{array} \right] [(\boldsymbol{\beta}_1^* \otimes \boldsymbol{\Lambda}^{\mathrm{H}})]_{:,r}+\tilde{\mathbf{N}}_{1}\nonumber
    \\&=\mathbf{\Theta}_{1}^{\mathrm{H}}\left[ \begin{array}{@{}c@{}}
    (\mathbf{A}_{M,1})_1^* \otimes (\mathbf{a}_{M}\left(\omega_{r},\mu_{r}\right))_1 \\
    (\mathbf{A}_{M,1})_2^* \otimes (\mathbf{a}_{M}\left(\omega_{r},\mu_{r}\right))_2 \\
    \vdots\\
    (\mathbf{A}_{M,1})_G^* \otimes (\mathbf{a}_{M}\left(\omega_{r},\mu_{r}\right))_G
    \end{array} \right] \alpha_r^*\boldsymbol{\beta}_1^*+\tilde{\mathbf{n}}_{1,r}\nonumber
    \\&=\mathbf{\Theta}_{1}^{\mathrm{H}} \mathbf{D}_{r,1}\alpha_r^*\boldsymbol{\beta}_1^*+\tilde{\mathbf{n}}_{1,r},
\label{eq:derive}
\end{align}
where $(\mathbf{A}_{M,1})_g^*$ and $(\mathbf{a}_{M}(\omega_{r},\mu_{r}))_g$ denote the $g$-th sub-blocks of $\mathbf{A}_{M,1}^*$ and $\mathbf{a}_{M}(\omega_{r},\mu_{r})$, respectively, and $\tilde{\mathbf{n}}_{1,r}$ is the $r$-th column of $\tilde{\mathbf{N}}_{1}$. 
This result is obtained by observing that the non-zero entries of $[(\boldsymbol{\beta}_1^* \otimes \boldsymbol{\Lambda}^{\mathrm{H}})]_{:,r}$ appear periodically at every $r$-th index within successive $L$-row blocks, forming the sequence $\{\alpha_r^*\beta_{1,j}^*\}_{j=1}^{J_1}$. Exploiting this specific block structure allows us to extract and concatenate the $r$-th column from each $L$-column block of $\mathbf{B}_1^{\mathrm{H}}$ to construct $\mathbf{D}_{r,1}$, which represents the block-Kronecker-structured RIS steering matrix associated with the $r$-th cascaded path. Consequently, this derivation yields $\mathbf{q}_{r,1}\triangleq \mathbf{D}_{r,1} \alpha_r^* \boldsymbol{\beta}_1^*$, exactly corresponding to the $r$-th column of $\mathbf{B}_1^{\mathrm{H}} (\boldsymbol{\beta}_1^* \otimes \boldsymbol{\Lambda}^{\mathrm{H}})$.

To estimate $\mathbf{q}_{r,1}$, Eq.~\eqref{eq:derive} can be rewritten in the virtual angular domain (VAD) as
\begin{align}
    \tilde{\mathbf{y}}_{1,r}&=\mathbf{\Theta}_{1}^{\mathrm{H}}\left[ \begin{array}{@{}c@{}}
    (\mathbf{A}_{1})_1^* \otimes (\mathbf{A}_{2})_1\\
    (\mathbf{A}_{1})_2^* \otimes (\mathbf{A}_{2})_2 \\
    \vdots \\
    (\mathbf{A}_{1})_G^* \otimes (\mathbf{A}_{2})_G
    \end{array} \right] \boldsymbol{b}_{1,r}+\tilde{\mathbf{N}}_{1} \nonumber\\
    &= \mathbf{\Theta}_1^{\mathrm{H}}\mathbf{D}\boldsymbol{b}_{1,r}+\tilde{\mathbf{N}}_{1}.
\label{eq:sparse}
\end{align}
Here, $({\mathbf{A}}_1)_g$ and $(\mathbf{A}_2)_g$ are the $g$-th row blocks of the corresponding overcomplete dictionary matrices, where $\mathbf{A}_{1} \in \mathbb{C}^{M\times D_1}$ and $\mathbf{A}_{2} \in \mathbb{C}^{M\times D_2}$ are overcomplete matrix ($D_1 \geq M, D_2 \geq M$) with resolutions $\frac{1}{D_1}$ and $\frac{1}{D_2}$, containing possible discretized values for $\{\mathbf{a}_{M,1}(\psi_{1,j},\nu_{1,j})\}_{j=1}^{J_1}$ and $\{\mathbf{a}_{M}(\omega_r,\mu_r)\}_{j=1}^{J_1}$ on the grid, respectively, and both are constructed as:
\begin{align}
    \mathbf{A} &= \mathbf{A}_v \otimes \mathbf{A}_h,\label{eq:dictionary1}\\\relax
    [\mathbf{A}_v]_{:,g_v} &= \mathbf{a}_{M_v}((-1+\frac{2}{D_v}g_v)\frac{d_{\mathrm{RIS}}}{\lambda_c}),\\\relax
    [\mathbf{A}_h]_{:,g_h} &= \mathbf{a}_{M_h}((-1+\frac{2}{D_h}g_h)\frac{d_{\mathrm{RIS}}}{\lambda_c}).
    \label{eq:dicitonary}
\end{align}
By concatenating all blocks to form the matrix $\mathbf{D}$, where $({\mathbf{A}}_{1})_g^* \otimes (\mathbf{A}_{2})_g$ has dimensions $\bar{M}^2 \times D_v D_h$, the equivalent dictionary matrix $\mathbf{\Theta}_1^{\mathrm{H}}\mathbf{D}\in\mathbb{C}^{\tau_1\times D_v D_h}$, and the sparse vector $\mathbf{b}_{1,r}$ with $J_1$ non-zero elements correspoing to the channel path gains $\{\alpha_r^*\beta_{1,j}^*\}_{j=1}^{J_1}$, we can perform estimation via sparse recovery algorithms, selecting $J_1$ atoms with the strongest correlations to estimate $\mathbf{q}_{r,1} = \mathbf{D}_{r,1} \alpha_r^* \boldsymbol{\beta}_1^*$, yielding the estimate $\mathbf{D}_{r,1}$ and $\alpha_r^* \boldsymbol{\beta}_1^*$.
To ensure optimal CS performance, the RIS phase shift matrix $\mathbf{\Theta}_1$ should be designed to ensure sufficient orthogonality between rows. A common method is to use the random Bernoulli matrix, i.e., generate $\mathbf{\Theta}_1$ by populating it with elements sampled uniformly at random from $\{-1, +1\}$, which is proved to be near-optimal $\cite{peng2022}$.
\label{subsec:corr_ls_est_other_paths}

$\textit{2) Estimation for other columns:}$
For the other columns, it is feasible to repeat the estimation procedure used for the reference colum, however, which would incur high computational complexity. In order to reduce the complexity, we can adopt the following approach.
For any other column $l \neq r$, we estimate $\mathbf{q}_{l,1} = \mathbf{D}_{l,1} \alpha_l^* \boldsymbol{\beta}_1^*$ and $\mathbf{y}_{1,l}=\mathbf{\Theta}_1^{\mathrm{H}}\mathbf{D}_{l,1}\alpha_l^*\boldsymbol{\beta}_1^*$ by exploiting its linear relationship with the reference column. Define the compensation matrix as $\Delta\mathbf{D}_{l,1}$, and note that the following relationship between $\mathbf{D}_{l,1}$ and $\mathbf{D}_{r,1}$ holds:
\begin{align}
\mathbf{D}_{l,1} = \Delta \mathbf{D}_{l,1} \mathbf{D}_{r,1},
\label{eq:linear eq}
\end{align}
where $\Delta \mathbf{D}_{l,1}$ is a block diagonal matrix, i.e. 
\begin{align}
    \Delta\mathbf{D}_{l,1} = \begin{bmatrix}
    (\Delta\mathbf{D}_{l,1})_1 & & & \\
    & (\Delta\mathbf{D}_{l,1})_2 & & \\
    & & \ddots & \\
    & & & (\Delta\mathbf{D}_{l,1})_G
    \end{bmatrix},
\end{align}
with its $g$-th block given by:
\begin{align}
(\Delta \mathbf{D}_{l,1})_g = \mathbf{I}_{\bar{M}} \otimes \operatorname{Diag}\left( (\mathbf{a}_{M}(\Delta \omega_l, \Delta \mu_l))_g \right),
\end{align}
and $\Delta \omega_l = \omega_l - \omega_r$, $\Delta \mu_l = \mu_l - \mu_r$. In addition, the gains are linearly related as follows:
\begin{align}
\alpha_l^* \boldsymbol{\beta}_1^* = \gamma_l^* \alpha_r^* \boldsymbol{\beta}_1^*,
\label{eq:gain_relation}
\end{align}
where $\gamma_l = \alpha_l / \alpha_r$. Denote $\operatorname{Diag}[(\mathbf{a}_M(\Delta\omega_l,\Delta\mu_l))_g]$ as $\operatorname{Diag}[(\mathbf{a}_{\Delta,l,g})]$, therefore, we can verify conclusion in Eq.~$\eqref{eq:linear eq}$ and Eq.~$\eqref{eq:gain_relation}$ in the following manner:
\begin{align}
&\gamma_l^* \Delta \mathbf{D}_{l,1} \mathbf{q}_{r,1} = \Delta\mathbf{D}_{l,1}\mathbf{D}_{r,1}\gamma_l^*\alpha_r^*\boldsymbol{\beta}_1^*\nonumber\\
&= \left[ \begin{array}{@{}c@{}}
    (\Delta\mathbf{D}_{l,1})_1(\mathbf{A}_{M,1})_1^* \otimes (\mathbf{a}_{M}\left(\omega_{r},\mu_{r}\right))_1 \\
    (\Delta\mathbf{D}_{l,1})_2(\mathbf{A}_{M,1})_2^* \otimes (\mathbf{a}_{M}\left(\omega_{r},\mu_{r}\right))_2 \\
    \vdots\\
    (\Delta\mathbf{D}_{l,1})_G(\mathbf{A}_{M,1})_G^* \otimes (\mathbf{a}_{M}\left(\omega_{r},\mu_{r}\right))_G
    \end{array} \right]\alpha_l^*\boldsymbol{\beta}_1^*\nonumber\\
&= \left[ \begin{array}{@{}c@{}}
    \mathbf{I}_{\bar{M}}(\mathbf{A}_{M,1})_1^* \otimes ((\operatorname{Diag}(\mathbf{a}_{\Delta,l,1}))(\mathbf{a}_{M}\left(\omega_{r},\mu_{r}\right))_1) \\
    \mathbf{I}_{\bar{M}}(\mathbf{A}_{M,1})_2^* \otimes ((\operatorname{Diag}(\mathbf{a}_{\Delta,l,2}))(\mathbf{a}_{M}\left(\omega_{r},\mu_{r}\right))_2) \\
    \vdots\\
    \mathbf{I}_{\bar{M}}(\mathbf{A}_{M,1})_G^* \otimes ((\operatorname{Diag}(\mathbf{a}_{\Delta,l,G}))(\mathbf{a}_{M}\left(\omega_{r},\mu_{r}\right))_G)
    \end{array} \right]\alpha_l^*\boldsymbol{\beta}_1^*\nonumber\\
&=\left[ \begin{array}{@{}c@{}}
    (\mathbf{A}_{M,1})_1^* \otimes (\mathbf{a}_{M}\left(\omega_{l},\mu_{l}\right))_1 \\
    (\mathbf{A}_{M,1})_2^* \otimes (\mathbf{a}_{M}\left(\omega_{l},\mu_{l}\right))_2 \\
    \vdots\\
    (\mathbf{A}_{M,1})_G^* \otimes (\mathbf{a}_{M}\left(\omega_{l},\mu_{l}\right))_G
    \end{array} \right]\alpha_l^*\boldsymbol{\beta}_1^* = \mathbf{q}_{l,1}.
\label{eq:linear trans}
\end{align}
This equality indicates that we can indirectly estimate $\mathbf{q}_{l,1}$ by estimating $\gamma_l^*$ and $\Delta\mathbf{D}_{l,1}$.
Substituting $\gamma_l^*\Delta\mathbf{D}_{l,1}$ into the expression for $\tilde{\mathbf{y}}_{1,l}$, we have the approximate model:
\begin{align}
\tilde{\mathbf{y}}_{1,l} = \boldsymbol{\Theta}_1^{\mathrm{H}} \widehat{\mathbf{q}}_{l,1} + \tilde{\mathbf{n}}_{1,l} = \gamma_l^* \boldsymbol{\Theta}_1^{\mathrm{H}} \Delta \mathbf{D}_{l,1} \widehat{\mathbf{q}}_{r,1} + \breve{\mathbf{n}}_{1,l},
\end{align}
where $\breve{\mathbf{n}}_{1,l} \triangleq \gamma_l^* \boldsymbol{\Theta}_1^{\mathrm{H}} \Delta \mathbf{D}_{l,1} \Delta{\mathbf{q}}_{r,1}+\tilde{\mathbf{n}}_{1,l}$ denotes the estimation residual of $\mathbf{q}_{r,1}$ and the original noise vector $\tilde{\mathbf{n}}_{1,l}$.
To estimate the unknown parameters $\Delta \omega_l$, $\Delta \mu_l$, and $\gamma_l^*$, we adopt the following two-step method:

\textbf{Step 1: Estimate the Rotation Parameters}

Via a 2-D maximal correlation search, the rotation parameters are estimated first:
\begin{align}
(\Delta \widehat{\omega}_l, \Delta \widehat{\mu}_l) = \arg \max_{\Delta \omega,\Delta \mu} \left| \tilde{\mathbf{y}}_{1,l}^{\mathrm{H}} \boldsymbol{\Theta}_1^{\mathrm{H}} \Delta \mathbf{D}(\Delta \omega, \Delta \mu) \widehat{\mathbf{q}}_{r,1} \right|,
\label{eq:deltaD}
\end{align}
where $\Delta \omega \in [-2\frac{d_{\mathrm{RIS}}}{\lambda_c}, 2\frac{d_{\mathrm{RIS}}}{\lambda_c}]$, $\Delta \mu \in [-2\frac{d_{\mathrm{RIS}}}{\lambda_c}, 2\frac{d_{\mathrm{RIS}}}{\lambda_c}]$.

\textbf{Step 2: Estimate the Gain Coefficient}

Let $\mathbf{c}_l \triangleq \boldsymbol{\Theta}_1^{\mathrm{H}} \Delta \mathbf{D}(\Delta \widehat{\omega}_l, \Delta \widehat{\mu}_l) \widehat{\mathbf{q}}_{r,1}$, then the LS estimate is:
\begin{align}
\widehat{\gamma}_l^* = (\mathbf{c}_l^{\mathrm{H}} \mathbf{c}_l)^{-1} \mathbf{c}_l^{\mathrm{H}} \tilde{\mathbf{y}}_{1,l}.
\label{eq:gamma}
\end{align}
Subsequently, the estimate for $\mathbf{q}_{l,1}$ is obtained as:
\begin{align}
\widehat{\mathbf{q}}_{l,1} = \widehat{\gamma}_l^* \Delta \mathbf{D}(\Delta \widehat{\omega}_l, \Delta \widehat{\mu}_l) \widehat{\mathbf{q}}_{r,1}.
\label{eq:ql}
\end{align}
Repeating the above steps for all $l \neq r$ yields the estimates for all columns [$\widehat{\mathbf{q}}_{1,1}, \ldots, \widehat{\mathbf{q}}_{L,1}$].
Finally, the estimated cascaded channel for the typical user can be expressed as:
\begin{align}
\widehat{\mathbf{G}}_1 = \widehat{\mathbf{A}}_N [\widehat{\mathbf{q}}_{1,1}, \ldots, \widehat{\mathbf{q}}_{L,1}]^{\mathrm{H}}.
\label{eq:G1}
\end{align}
The computational complexity of this estimation method are significantly lower than those of the method that applies OMP separately to each path, which will later be illustrated in Section~\ref{sec:pilot_and_complexity}. The complete algorithm is shown in Algorithm~$\ref{alg:fullcsi_estimation}$. 
\begin{algorithm}[H]
\caption{Estimation of full CSI for Typical User}
\label{alg:fullcsi_estimation}
\begin{algorithmic}[1]
\Require $\widehat{L}$ and $\widehat{\mathbf{A}}_N$ from Algorithm~$\ref{alg:aoa_estimation}$, $\mathbf{Y}_{1}$
\State Calculate the equivalent measurement matrix: $\tilde{\mathbf{Y}}_{k} = \left(\frac{1}{N\sqrt{p}}\mathbf{A}_N^{\mathrm{H}} \mathbf{Y}_{k}\right)^{\mathrm{H}}$.
\State Calculate equivalent dictionary $\mathbf{\Theta}_1^{\mathrm{H}}\mathbf{D}$.
\State Estimate the reference column $\mathbf{q}_{r,1}$ via solving sparse recovery problem in Eq.~$\eqref{eq:sparse}$
\For{$l=1$ to $\widehat{L}$, $l\neq r$}
    \State Calculate $\Delta\mathbf{D}_{l,1}$ via Eq.~$\eqref{eq:deltaD}$.
    \State Calculate $\gamma_l^*$ via Eq.~$\eqref{eq:gamma}$.
    \State Calculate $\widehat{\mathbf{q}}_{l,1}$, according to Eq.~$\eqref{eq:ql}$.
    \State Estimate $\widehat{\mathbf{G}}_1$ via Eq.~\eqref{eq:G1}.
\EndFor
\Ensure $\widehat{\mathbf{G}}_1=\widehat{\mathbf{A}}_N [\widehat{\mathbf{q}}_{1,1}, \ldots, \widehat{\mathbf{q}}_{L,1}]^{\mathrm{H}}$.
\end{algorithmic}
\end{algorithm}

\subsection{Stage III: Estimation for Other Users}
\label{subsec:est_other_users}
At this stage, we aim to leverage the estimation results from Stage I and Stage II to separate the user's cascaded channel into a common part and a user-specific part, thereby reducing pilot overhead while preserving high estimation accuracy.
\subsubsection{Construction of the Common Part}
Since all users share the common AoA steering matrix $\mathbf{A}_N$, after obtaining $\widehat{\mathbf{A}}_N$ from user 1, it is reused for the remaining users. 
However, the complete common part $\mathbf{H}$ cannot be directly recovered from $\widehat{\mathbf{G}}_1$. In Stage II, we can acquire the cascaded angle matrices and gains (i.e., $\mathbf{D}_{l,1}$ and $\alpha_l^*\boldsymbol{\beta}_1^*$), but these quantities are still in a cascaded form. Moreover, under the \textit{group-connected} BD-RIS architecture, the cascaded angle information in $\mathbf{D}_{l,1}$ is not directly extractable due to the information loss from the grouped structure, so only limited information can be reused.
Therefore, a equivalent common part $\mathbf{H}_s$ is constructed from $\widehat{\mathbf{G}}_1$ to preserve the shared information required in Stage III.

Based on the estimates of parameters $\Delta\mathbf{D}_{l,1}$ and $\gamma_l^*$ for the typical user from Eq.~$\eqref{eq:deltaD}$ and Eq.~$\eqref{eq:gamma}$, we can obtain the following information:
$\mathbf{a}_M(\Delta\widehat{\omega_l},\Delta\widehat{\mu_l})$
and
$\widehat{\boldsymbol{\beta}_1^*\alpha_l^*} = \gamma_l^*\widehat{\boldsymbol{\beta}_1^*\alpha_r^*}$.
Utilizing this known information, we obtain:
\begin{align}
\boldsymbol{\Lambda} &= \operatorname{Diag}(\alpha_1,\alpha_2,\ldots,\alpha_L)=\alpha_r\operatorname{Diag}(\gamma_1,\gamma_2,\ldots,\gamma_L) \nonumber \\
&= \frac{1}{\bar{\beta}}\bar{\beta}\alpha_r\operatorname{Diag}(\gamma_1,\gamma_2,\ldots,\gamma_L) = \frac{1}{\bar{\beta}}\boldsymbol{\Lambda}_s,
\end{align}
and
\begin{align}
\mathbf{A}_M = \operatorname{Diag}\{\mathbf{a}_M(\omega_r,\mu_r)\}\mathbf{A}_{\Delta M}.
\end{align}
where $\mathbf{A}_{\Delta M} \triangleq [\mathbf{a}_M(\Delta\omega_1,\Delta\mu_1),\ldots,\mathbf{a}_M(\Delta\omega_L,\Delta\mu_L)]$ can be readily estimated via observing the structure of Eq.~$\eqref{eq:deltaD}$.
Then, $\mathbf{H}$ can be rewritten as:
\begin{align}
\mathbf{H} &= \frac{1}{\bar{\beta}}\mathbf{A}_N\mathbf{\Lambda}_s\mathbf{A}_{\Delta M}^{\mathrm{H}}\operatorname{Diag}\{\mathbf{a}_M(-\omega_r,-\mu_r)\}\nonumber\\&=\frac{1}{\bar{\beta}}\mathbf{H_s}\operatorname{Diag}\{\mathbf{a}_M(-\omega_r,-\mu_r)\},
\end{align}
where $ \mathbf{H}_s\triangleq \mathbf{A}_N\mathbf{\Lambda}_s\mathbf{A}_{\Delta M}^{\mathrm{H}}$ is defined as the common part of cascaded channels. 
\subsubsection{Estimation of the User-Specific Part}
For simplicity, in the following, we denote $\mathbf{a}_M(-\omega_r,-\mu_r)$ as $\mathbf{a}_M$, and the previous channel model Eq.~\eqref{eq:cascaded_group_sum} of the user $k$ can be reformulated as:
\begin{align}
&\mathbf{h}_{\mathrm{BD},k}=\mathbf{H}\mathbf{\Phi}\mathbf{h_{k}}=\frac{1}{\bar{\beta}}\mathbf{H_s}\operatorname{Diag} \{\mathbf{a}_M(-\omega_r,-\mu_r)\}\mathbf{\Phi}\mathbf{h_{k}} \nonumber \\
&= \frac{1}{\bar{\beta}}\mathbf{H}_s\left[ \begin{array}{@{}c@{}}
    [\mathbf{a}_M]_1 [\mathbf{\Phi}\mathbf{h}_k]_1 \\\relax
    [\mathbf{a}_M]_2 [\mathbf{\Phi}\mathbf{h}_k]_2\\
    \vdots \\\relax
    [\mathbf{a}_M]_M [\mathbf{\Phi}\mathbf{h}_k]_M
    \end{array} \right] 
= \frac{1}{\bar{\beta}}\mathbf{H}_s\left[ \begin{array}{@{}c@{}}
    [\mathbf{\Phi}]_{1,:}[\mathbf{a}_M]_1\mathbf{h}_k \\\relax
    [\mathbf{\Phi}]_{2,:}[\mathbf{a}_M]_2\mathbf{h}_k\\
    \vdots \\\relax
    [\mathbf{\Phi}]_{M,:}[\mathbf{a}_M]_M\mathbf{h}_k
    \end{array} \right]\nonumber\\
&=  \frac{1}{\bar{\beta}}\mathbf{H}_s\begin{bmatrix}
    \lbrack\mathbf{\Phi}\rbrack_{1,\text{:}} & & & \\
    & \lbrack\mathbf{\Phi}\rbrack_{2,\text{:}} & & \\
    & & \ddots & \\
    & & & \lbrack\mathbf{\Phi}\rbrack_{M,\text{:}}
    \end{bmatrix}(\mathbf{a}_M \otimes \mathbf{h}_k)
    \nonumber\\
&= \mathbf{H}_s \mathbf{P} (\mathbf{a}_M \otimes \mathbf{h}_k),
\label{eq:multi}
\end{align}
where $\mathbf{P}=\operatorname{blkdiag}\{\lbrack\mathbf{\Phi}\rbrack_{1,:},\ldots,\lbrack\mathbf{\Phi}\rbrack_{M,:}\}$ and $(\mathbf{a}_M \otimes \mathbf{h}_k)$ denotes the user-specific part for user $k$. Exploiting the array response structure of $\mathbf{a}_M$ and $\mathbf{h}_k$, the received signal $\mathbf{y}_k(t)$ of time slot $t$ can be sparsely represented as
\begin{align}
\mathbf{y}_k(t) &= \mathbf{H}_s \mathbf{P} (\mathbf{a}_M \otimes \mathbf{h}_k)+\mathbf{n}_t^{\mathrm{noise}}\nonumber\\&=\mathbf{H}_s \mathbf{P}(\tilde{\mathbf{A}}_1\boldsymbol{\gamma})\otimes(\tilde{\mathbf{A}}_2\boldsymbol{\lambda}_k)+\mathbf{n}_t^{\mathrm{noise}}
\nonumber\\&=\mathbf{H}_s \mathbf{P}(\tilde{\mathbf{A}}_1\otimes\tilde{\mathbf{A}}_2)(\dfrac{1}{\bar{\beta}}\boldsymbol{\gamma}\otimes\boldsymbol{\lambda}_k)+\mathbf{n}_t^{\mathrm{noise}},
\label{eq:multi_sparse_front}
\end{align}
where $\mathbf{a}_M = \tilde{\mathbf{A}}_1 \boldsymbol{\gamma}$, $\mathbf{h}_{k} = \mathbf{A}_{M,k} \boldsymbol{\beta}_k = \tilde{\mathbf{A}}_{2} \boldsymbol{\lambda}_k$, $\mathbf{n}_t^{\mathrm{noise}}$ is the corresponding noise. $\tilde{\mathbf{A}}_1 \in \mathbb{C}^{M\times P_1}$ and $\tilde{\mathbf{A}}_{2} \in \mathbb{C}^{M\times P_2}$ are known overcomplete dictionary matrices with similar structures in Eq.~$\eqref{eq:dictionary1}$--$\eqref{eq:dicitonary}$. $\boldsymbol{\gamma} \in \mathbb{C}^{P_1\times 1}$ and $\boldsymbol{\lambda}_k \in \mathbb{C}^{P_2\times 1}$ are sparse vectors to be estimated, with the former being 1-sparse.

Given that $\mathbf{H}_s \mathbf{P}(\tilde{\mathbf{A}}_1\otimes\tilde{\mathbf{A}}_2)$ is specified as the equivalent cascaded dictionary, and $(\mathbf{a}_M \otimes \mathbf{h}_k)$ is the sparse vector, it can be typically solved by Eq.~$\eqref{eq:multi_sparse_front}$ via sparse recovery. However, the sparsity of matrix $\mathbf{P}$ depends on the number of groups $G$, with the first $\bar{M}$ elements of every $M$ elements in $\{\lbrack\mathbf{\Phi}\rbrack_{i,:}\}_{i=1}^M$ being non-zero. Clearly, $\mathbf{P}$ constructed directly in this way is overly sparse, which degrades the orthogonality of the equivalent dictionary, causing OMP to fail in accurate estimation. Considering the grouping structure of the BD-RIS, let $(\mathbf{a}_M)_g \in \mathbb{C}^{\bar{M}\times 1}$ be the $g$-th group of $\mathbf{a}_M$ and define $\operatorname{Diag}\{(\mathbf{a}_M)_g\}$ as $\mathbf{A}_{g}^{d}$, we construct the grouped expression:
\begin{align}
&\mathbf{h}_{\mathrm{BD},k}=\frac{1}{\bar{\beta}}\mathbf{H}_s
\begin{bmatrix}
\mathbf{A}_{1}^{d} & & & \\
& \mathbf{A}_{2}^{d} & & \\
& & \ddots & \\
& & & \mathbf{A}_{G}^{d}
\end{bmatrix}\begin{bmatrix}
\overline{\mathbf{\Phi}}_{1}\mathbf{h}_{k,1}\\
\overline{\mathbf{\Phi}}_{2}\mathbf{h}_{k,2}\\
\vdots \\
\overline{\mathbf{\Phi}}_{G}\mathbf{h}_{k,G}
\end{bmatrix} \nonumber \\
&=\frac{1}{\bar{\beta}}\mathbf{H_s}\begin{bmatrix}
\mathbf{A}_{1}^{d}\overline{\mathbf{\Phi}}_{1}\mathbf{h}_{k,1} \\
\mathbf{A}_{2}^{d}\overline{\mathbf{\Phi}}_{2}\mathbf{h}_{k,2}\\
\vdots \\
\mathbf{A}_{G}^{d}\overline{\mathbf{\Phi}}_{G}\mathbf{h}_{k,G}
\end{bmatrix} = \frac{1}{\bar{\beta}}\mathbf{H_s}\begin{bmatrix}
\mathbf{P}_1((\mathbf{a}_M)_1 \otimes \mathbf{h}_{k,1}) \\
\mathbf{P}_2((\mathbf{a}_M)_2 \otimes \mathbf{h}_{k,2})\\
\vdots \\
\mathbf{P}_G((\mathbf{a}_M)_G \otimes \mathbf{h}_{k,G})
\end{bmatrix},
\end{align}
where $\{\mathbf{P}_g\}_{g=1}^G = \operatorname{blkdiag}\{\lbrack\mathbf{\bar{\Phi}_g}\rbrack_{1,:},\lbrack\mathbf{\bar{\Phi}_g}\rbrack_{2,:},\ldots,\lbrack\mathbf{\bar{\Phi}_g}\rbrack_{\bar{M},:}\}$ has a similar structure with $\mathbf{P}$.
Thus, an underdetermined problem is transformed into a solvable sparse recovery problem. Our objective is to solve for the user-specific part $\{((\mathbf{a}_M)_g \otimes \mathbf{h}_{k,g})\}_{g=1}^G$.
With the grouped structure, the received signal $\mathbf{y}_k(t)$ of time slot $t$ can be sparsely represented as
\begin{align}
\mathbf{y}_k(t) =\mathbf{H}_s\begin{bmatrix}
\mathbf{P}_1((\tilde{\mathbf{A}}_{1})_1\otimes(\tilde{\mathbf{A}}_{2})_1) \\
\mathbf{P}_2((\tilde{\mathbf{A}}_{1})_2\otimes(\tilde{\mathbf{A}}_{2})_2)  \\
\vdots \\
\mathbf{P}_G((\tilde{\mathbf{A}}_{1})_G\otimes(\tilde{\mathbf{A}}_{2})_G)
\end{bmatrix}\frac{1}{\bar{\beta}}\boldsymbol{\gamma} \otimes \boldsymbol{\lambda}_k+\mathbf{n}_t^{\mathrm{noise}},
\label{eq:multi_sparse}
\end{align}
where $(\mathbf{a}_M)_g = (\tilde{\mathbf{A}}_1)_g \boldsymbol{\gamma}$, $\mathbf{h}_{k,g} = (\mathbf{A}_{M,k})_g \boldsymbol{\beta}_k = (\tilde{\mathbf{A}}_{2})_g \boldsymbol{\lambda}_k$ are the corresponding grouped sparse representations.

With an optimized group sparse representation, directly solving Eq.~\eqref{eq:multi_sparse} over a $D_1 D_2$-dimensional equivalent dictionary is still suboptimal with the following reasons. The grouping matrix $\mathbf{P}_g$ aggravates column correlation in the equivalent dictionary, while conventional estimators fail to exploit the 1-sparse structural prior of $\boldsymbol{\gamma}$. To overcome these limitations, we propose the HBOMP algorithm, which decomposes the recovery into a two-step procedure to constrain the search dimension, thus successfully addressing these issues. 

The algorithmic implementation of HBOMP is established upon the block configurations. Due to the 1-sparse nature of $\boldsymbol{\gamma}$, the Kronecker product $\boldsymbol{\gamma} \otimes \boldsymbol{\lambda}_k$ strictly exhibits an intra-block sparsity pattern. Specifically, if the non-zero entry of $\boldsymbol{\gamma}$ occurs at index $j^* \in \{1,\dots,P_{1}\}$, only the $j^*$-th block of size $P_2$ in $\boldsymbol{\gamma} \otimes \boldsymbol{\lambda}_k$ contains non-zero elements. Let $\mathbf{a}_{1,g}^{(j)}$ denote the $j$-th column of $(\mathbf{A}_{1})_g$. For any hypothesized index $j$, the dictionary matrix for group $g$ at time slot $t$ becomes $\mathbf{\Phi}_{t,g}^{(j)} = \mathbf{P}_g ( \mathbf{a}_{1,g}^{(j)} \otimes (\tilde{\mathbf{A}}_2)_g )$. Stacking these across all $G$ groups, we can define the time-slot dictionary block corresponding to the $j$-th atom of $\boldsymbol{\gamma}$ as:
\begin{align}
\mathbf{\Phi}_t^{(j)} = \begin{bmatrix} 
\mathbf{P}_1 ( \mathbf{a}_{1,1}^{(j)} \otimes (\tilde{\mathbf{A}}_2)_1 ) \\ 
\vdots \\ 
\mathbf{P}_G ( \mathbf{a}_{1,G}^{(j)} \otimes (\tilde{\mathbf{A}}_2)_G )
\end{bmatrix}=\begin{bmatrix} 
\mathbf{\Phi}_{t,1}^{(j)} \\ 
\vdots \\ 
\mathbf{\Phi}_{t,G}^{(j)} 
\end{bmatrix}.
\end{align}
Vectorizing the received signal matrix $\mathbf{Y}_k$ across $\tau_k$ time slots into $\mathbf{y}_k^{s}$ and eliminating the common AoA steering matrix at the BS $\mathbf{A}_N$, we obtain a highly structured linear expression parameterized by the active block:
\begin{align}
    \mathbf{y}_k^{s} =
\begin{bmatrix}
    \mathbf{H}_c\mathbf{\Phi}_1^{(j^*)}\\
    \vdots  \\
    \mathbf{H}_c\mathbf{\Phi}_{\tau_k}^{(j^*)}\\
\end{bmatrix}\frac{1}{\bar{\beta}}\boldsymbol{\lambda}_k+\mathbf{n}_{stack}
\triangleq \mathbf{\Psi}^{(j^*)}\tilde{\boldsymbol{\lambda}}_k +\mathbf{n}_{stack},
\end{align}
where $\mathbf{H}_c \triangleq \frac{1}{N\sqrt{p}}\widehat{\mathbf{A}}_N^{\mathrm{H}}\mathbf{H}_s$ represents the equivalent common channel matrix, $\mathbf{\Psi}^{(j^*)} \in \mathbb{C}^{L\tau_k \times P_2}$ is the equivalent dictionary block for the index $j^*$, and $\tilde{\boldsymbol{\lambda}}_k = \frac{1}{\bar{\beta}}\boldsymbol{\lambda}_k$. 

Based on this decomposed structure, the global block dictionary is defined as $\mathbf{\Psi} = [\mathbf{\Psi}^{(1)}, \mathbf{\Psi}^{(2)}, \ldots, \mathbf{\Psi}^{(P_1)}] \in \mathbb{C}^{L\tau_k \times P_1 P_2}$. Initially, by evaluating the $\ell_\infty$-norm of the correlations for all $P_1$ blocks, the algorithm isolates the active macro-block $j^*$ associated with the mutual BS-RIS channels. After extracting this active block $\mathbf{\Psi}^{(j^*)}$, we apply a standard OMP with a residual threshold to recover the sparse vector $\tilde{\boldsymbol{\lambda}}_k$. The detailed steps are summarized in Algorithm~\ref{alg:bomp}.

\begin{algorithm}
\caption{HBOMP based estimation of $\mathbf{G}_k, 2 \leq k \leq K$}
\label{alg:bomp}
\textbf{Input:} $\mathbf{Y}_k, 2 \leq k \leq K$.     
\begin{algorithmic}[1]
\State Return parameters from Stage I.
\State Calculate $\mathbf{y}_k^{s}=\frac{1}{N\sqrt{p}}\operatorname{vec}(\mathbf{Y}_k)$.
\State Calculate the common channel $\mathbf{H}_c$.
\State Construct the global dictionary blocks $\mathbf{\Psi}^{(j)}$ for $j=1,\ldots,P_1$.
\State \textbf{Macro-Block Identification:}
\State Get normalized blocks $\tilde{\mathbf{\Psi}}^{(j)}$ via column scaling on $\mathbf{\Psi}^{(j)}$.
\State Find active block $j^* = \arg\max\limits_{j=1,\dots,P_1} \| (\tilde{\mathbf{\Psi}}^{(j)})^{\mathrm{H}} \mathbf{y}_k^{s} \|_\infty$.
\State Set the active dictionary $\bar{\mathbf{\Psi}} = \tilde{\mathbf{\Psi}}^{(j^*)}$.
\State \textbf{Intra-Block Sparse Recovery:}
\State Initialize residual $\mathbf{r}_0 = \mathbf{y}_k^{s}$, support set $\Omega_0 = \emptyset$, index $i=1$.
\State \textbf{Repeat}
\State Find atom $idx = \arg\max\limits_{m=1,\dots,P_2} |(\bar{\mathbf{\Psi}})_{(:,m)}^{\mathrm{H}} \mathbf{r}_{i-1}|$.
\State Update support: $\Omega_i = \Omega_{i-1} \cup \{idx\}$.
\State LS solution: $\mathbf{b}_i = (\bar{\mathbf{\Psi}}(:,\Omega_i)^{\mathrm{H}} \bar{\mathbf{\Psi}}(:,\Omega_i))^{-1} \bar{\mathbf{\Psi}}(:,\Omega_i)^{\mathrm{H}} \mathbf{y}_k^{s}$.
\State Update residual: $\mathbf{r}_i = \mathbf{y}_k^{s} - \bar{\mathbf{\Psi}}(:,\Omega_i) \mathbf{b}_i$.
\State $i = i + 1$.
\State \textbf{Until} $\|\mathbf{r}_{i-1}\|_2 \leq \text{threshold}$
\State Obtain the estimates:
\State $\tilde{\boldsymbol{\lambda}}_k(\Omega_{i-1}) = \mathbf{b}_{i-1}$.
\State Set $\boldsymbol{\gamma}$ as a 1-sparse vector with $1$ at index $j^*$.
\State Reconstruct target: $\frac{1}{\bar{\beta}}(\mathbf{a}_M)_g \otimes \mathbf{h}_{k,g} = (\tilde{\mathbf{A}}_1)_g\boldsymbol{\gamma} \otimes (\tilde{\mathbf{A}}_2)_g\tilde{\boldsymbol{\lambda}}_k$.
\State Estimate $\mathbf{G}_k$ from $\{(\mathbf{a}_M)_g \otimes \mathbf{h}_{k,g}\}_{g=1}^G$ according to Eq.~\eqref{eq:gk} and Eq.~\eqref{eq:gk_final}.     
\Ensure $\widehat{\mathbf{G}}_k$, $2 \leq k \leq K$.
\end{algorithmic}
\end{algorithm}
\subsubsection{Recover the cascaded channel}
The OMP algorithm utilized for the sparse recovery problem directly identifies both the dictionary atoms and their corresponding coefficients. Thus, from the estimated target $\frac{1}{\bar{\beta}}((\mathbf{a}_M)_g \otimes \mathbf{h}_{k,g})$, we can explicitly extract its corresponding angular composite term $\left( (\mathbf{a}_M(-\omega_r,-\mu_r))_g \otimes (\mathbf{A}_{M,k})_g \right)$ and the associated equivalent path gain $\frac{1}{\bar{\beta}}\boldsymbol{\beta}_k$, which satisfy
\begin{align}
&\frac{1}{\bar{\beta}}((\mathbf{a}_M)_g \otimes \mathbf{h}_{k,g}) 
= (\mathbf{a}_M(-\omega_r,-\mu_r))_g \otimes ( (\mathbf{A}_{M,k})_g\frac{1}{\bar{\beta}}\boldsymbol{\beta}_k )\nonumber\\
&= \left( (\mathbf{a}_M(-\omega_r,-\mu_r))_g \otimes (\mathbf{A}_{M,k})_g \right)\frac{1}{\bar{\beta}}\boldsymbol{\beta}_k.
\end{align}
To formulate the components of the cascaded channel $\mathbf{G}_k$, we swap the Kronecker product order of the extracted angular composite term $\left( (\mathbf{a}_M(-\omega_r,-\mu_r))_g \otimes (\mathbf{A}_{M,k})_g \right)$ through matrix reshaping and reordering. Taking the complex conjugate of this permuted angle alongside its associated path gain $\frac{1}{\bar{\beta}}\boldsymbol{\beta}_k$ yields the desired cascaded component:
\begin{align}
\bar{\mathbf{g}}_k\triangleq\left( (\mathbf{A}_{M,k})_g^* \otimes (\mathbf{a}_M(\omega_r,\mu_r))_g \right)\frac{1}{\bar{\beta}^*}\boldsymbol{\beta}_k^*,
\end{align}
where we denote the result $((\mathbf{A}_{M,k})_g^* \otimes (\mathbf{a}_M(\omega_r,\mu_r))_g)\frac{1}{\bar{\beta}^*}\boldsymbol{\beta}_k^*$ as a temporary cascaded channel $\bar{\mathbf{g}}_k$.
Since $\mathbf{\Lambda}_s(l,l)=\bar{\beta}\alpha_l$, we obtain $\alpha_l^*\boldsymbol{\beta}_k^* = \frac{1}{\bar{\beta}^*}\boldsymbol{\beta}_k^*\mathbf{\Lambda}_s^*(l,l)$. Using the known $\mathbf{\Lambda}_s$, $\mathbf{a}_M(\Delta\widehat{\omega}_l,\Delta\widehat{\mu}_l)$, and the conclusion in Eq.~$\eqref{eq:linear trans}$, we have:
\begin{align}
\widehat{\mathbf{q}}_{l,k} &= (\mathbf{A}_{M,k})_g^* \otimes (\mathbf{a}_{M}\left(\omega_{l},\mu_{l}\right))_g\alpha_l^*\boldsymbol{\beta}_k^* \nonumber \\
&= (\mathbf{I}_{\bar{M}} \otimes \operatorname{Diag}( \mathbf{a}_{M}(\Delta \widehat{\omega}_l, \Delta \widehat{\mu}_l) )_g)\bar{\mathbf{g}}_k\mathbf{\Lambda}_s^*  (l,l).
\label{eq:gk}
\end{align}
All parameters are well estimated, solving Eq.~$\eqref{eq:gk}$ recovers $\mathbf{q}_{l,k}$.
Stacking $\widehat{\mathbf{q}}_{l,k}$ for all $l$ and subsequently we can readily recover $\mathbf{G}_k$:
\begin{align}
\widehat{\mathbf{G}}_k = \sqrt{p} \widehat{\mathbf{A}}_N [\widehat{\mathbf{q}}_{1,k},\widehat{\mathbf{q}}_{2,k},\ldots,\widehat{\mathbf{q}}_{L,k}]^{\mathrm{H}}.
\label{eq:gk_final}
\end{align}
 
\section{Pilot Overhead and Computation Complexity Analysis}
\label{sec:pilot_and_complexity}
In this section, we analyze the pilot overhead and computation complexity of our proposed protocol, with comparison to other baseline algorithms in Table~\ref{tab:comparison_with_lines}.
For convenience, we assume $J_1 = J_2 = \ldots = J_K = J$.
\subsection{Pilot Overhead}
\label{subsec:pilot_overhead}
To find an $l$-sparse complex signal (vector) of dimension $n$, the required number of measurements $m$ is on the order of $\mathcal{O}(l\operatorname{log}(n))$ \cite{candes2006}. For the pilot overhead in Stage I and II (typical user), the equivalent sensing matrix $\mathbf{D} = \mathbf{\Theta}_{k}^{\mathrm{H}}[(\mathbf{A}_{M,k})_1^{\mathrm{T}} \otimes (\mathbf{A}_M^{\mathrm{H}})_1, \ldots, (\mathbf{A}_{M,k})_G^{\mathrm{T}} \otimes (\mathbf{A}_M^{\mathrm{H}})_G]^{\mathrm{H}}$ has dimensions $\tau_1 \times D_1D_2$, where $D_1 \geq M$ and $D_2 \geq M$, and the corresponding sparsity level is $J$. Therefore, the pilot overhead for user 1 should satisfy
$\tau_1 \geq \mathcal{O}(J\operatorname{log}(D_1D_2)) \geq \mathcal{O}(J\operatorname{log}(M^2))$.
In simulations, the dictionary resolution is typically set to quadruple the dimension, i.e., $D_1 = 4M, D_2=4M$, leading to $\tau_1 \geq \mathcal{O}(J\operatorname{log}(16M^2))$.

For other users, the equivalent HBOMP sensing matrix evaluates a block dictionary $\mathbf{\Psi} \in \mathbb{C}^{L\tau_k \times P_1 P_2}$. According to model-based CS~\cite{baraniuk2010}, the theoretical measurement bound for recovering a structured sparse signal is $\mathcal{O}(J + \log|\mathcal{C}|)$, where $\mathcal{C}$ denotes the set of permissible sparsity patterns. In our HBOMP architecture, identifying a valid sparsity pattern operates in two stages: locating the exact active macro-block among the $P_1$ candidates, and subsequently picking $J$ active paths out of the $P_2$ angular grid points within that specific macro-block. The number of ways to distribute these $J$ paths across the $P_2$-dimensional block corresponds to the binomial coefficient $\binom{P_2}{J}$. Consequently, the total number of configurations is $|\mathcal{C}| = P_1 \binom{P_2}{J}$. Substituting this into the bound yields $L\tau_k \geq \mathcal{O}\big(J + \log(P_1 \binom{P_2}{J})\big)$. By applying the asymptotic combinatorial upper bound $\log \binom{P_2}{J} = \mathcal{O}(J \log(P_2))$, the required measurement dimension rigorously simplifies to $L\tau_k \geq \mathcal{O}(\log(P_1) + J \log(P_2))$. Recalling the dimensional equivalence $P_1 = 4M = P_2$, this structural bound directly reduces to $\mathcal{O}((J+1)\log(4M))$. Therefore, the pilot overhead allocated for user $k$ reduces to
$\tau_k \geq \mathcal{O}\left( {(J+1)\operatorname{log}(4M)}/{L} \right).$
Therefore, the total pilot overhead is $\tau_{total} \geq \mathcal{O}(J\operatorname{log}(16M^2)+(K-1)(J+1)\operatorname{log}(4M)/L)$.
\begin{table*}[t]
\centering
\caption{Comparison of Pilot Overhead and Computation Complexity}
\label{tab:comparison_with_lines}
\begin{tabular}{|c|l|c|c|}
\hline
\multicolumn{2}{|c|}{\textbf{Algorithm}} & \textbf{Pilot Overhead} & \textbf{Complexity} \\
\hline
\multicolumn{2}{|c|}{\makecell{Proposed protocol\\}} 
 & \makecell{$\mathcal{O}(J\operatorname{log}(16M^2)+$\\$(K-1)(J+1)\operatorname{log}(4M)/L)$ }& \makecell{$\mathcal{O}( ( gL N + d(L-1) M^2/G )\tau_1$\\$+(K-1)(16M^4/G + L(16M^2 + 4JM))\tau_k)$} \\
\cline{2-3}
\hline
\multicolumn{2}{|c|}{Direct-OMP algorithm} 
& $\mathcal{O}(KJL^2\operatorname{log}(64M^2N))$ & $\mathcal{O}(K(N^3M^3+64JL^2N^2M^2)\tau_k)$ \\
\hline
\multicolumn{2}{|c|}{SBL-based algorithm} 
& - - & $\mathcal{O}(K(M^6+M^4N+(M^4+M^2N)\tau_k))$ \\
\hline
\end{tabular}
\end{table*}   
\subsection{Computation Complexity}
\label{subsec:computation_complexity}

In this subsection, we analyze the computational complexity of the proposed three-stage channel estimation algorithm. 
\subsubsection{Stage I: BS AoA Estimation}
Stage I consists of the 2D-DFT and angle rotation. The complexity of Stage I is dominated by the calculation of angle rotation in $\cref{alg:aoa_estimation}$.

For each of the ${L}$ estimated paths, the algorithm performs two separate 1D searches,i.e., Eqs. (30) and (31), over grids of size $g_1$ and $g_2$, respectively. Each search involves computing a correlation with complexity $\mathcal{O}(N \tau_1)$. Therefore, the total complexity for angle rotation is $\mathcal{O}(L (g_1 + g_2) N \tau_1)$. 

\subsubsection{Stage II: Full CSI Estimation for the Typical User}
The complexity of Stage II involves estimating the reference column via OMP and the remaining columns via correlation and LS estimation. The dominant cost is the correlation search while estimating other columns in $\cref{alg:fullcsi_estimation}$.

For the remaining ${L}-1$ paths, the algorithm performs a 2D correlation search, Eq. (48) and an LS estimation, Eq. (49). The correlation search over a 2D grid of size $d_1d_2$ involves computing $\mathbf{\Theta}_1^{\mathrm{H}} \Delta \mathbf{D}_{l,1} \widehat{\mathbf{q}}_{r,1}$ for each grid point, with complexity $\mathcal{O}((d_1d_2 M^2/ G) \tau_1)$. The subsequent LS estimation for the scalar $\widehat{\gamma}_l^*$ has complexity $\mathcal{O}(\tau_1)$. Thus, over ${L}$ searches, the complexity for all other columns is $\mathcal{O}((({L}-1) d_1d_2 M^2/ G )\tau_1)$.

\subsubsection{Stage III: CSI Estimation for Other Users via HBOMP}
The complexity for estimating user $k$ is dominated by sparse recovery in $\cref{alg:bomp}$. Due to the grouped BD-RIS layout, generating the equivalent dictionary blocks scales down to $\mathcal{O}((16M^4/G) \tau_k)$ operations. According to~\cite{venugopal2017}, recovering an $S$-sparse signal from an $M \times N$ dictionary via OMP requires $\mathcal{O}(S M N)$ operations. Leveraging this property, identifying the active block from $16M^2$ candidates and extracting the $J$-sparse vector from $4M$ atoms via OMP has the complexity of $\mathcal{O}(L\tau_k (16M^2 + 4JM))$.
Thus, the total complexity for estimating the remaining $K-1$ users in Stage III is bounded by $\mathcal{O}((K-1) [ 16M^4/G + L(16M^2 + 4JM)]\tau_k)$.

\subsubsection{Overall Complexity}
Considering only the dominant terms, let $g \triangleq g_1+g_2$ and $d \triangleq d_1d_2$, the overall computation complexity is given by $\mathcal{O}( (gLN + d(L-1) M^2/G)\tau_1 + (K-1)( 16M^4/G + L(16M^2 + 4JM))\tau_k )$.
Notably, as the grouping number $G$ increases, the overall computational complexity reduces proportionally.

As shown in Table~\ref{tab:comparison_with_lines}, in the BD-RIS scenario, the proposed three-stage protocol significantly reduces both the pilot overhead and computational complexity compared to baseline algorithms. The subsequent simulation results will further confirm this fact.

\section{Simulation Results}
\label{sec:simulation_results}

This section evaluates the performance of the proposed three-stage channel estimation protocol for a BD-RIS-aided MU-MISO system via simulations. The BS is positioned at $100 \, \text{m} \times (-1,0,0)$, the RIS is deployed at the origin $(0,0,0)$, and the users are uniformly distributed within a sphere of radius $1 \, \text{m}$, centered at $10 \, \text{m} \times (1,0,0)$. The uplink carrier frequency is $f = 28 \, \text{GHz}$. The BS employs a UPA of dimensions $N = N_h \times N_v = 8 \times 8 = 64$, and the BD-RIS comprises a passive reflecting UPA with $M = M_h \times M_v = 6 \times 6 = 36$ elements. The $M$ elements are connected to an $M$-port network and are uniformly divided into $G$ groups. The antenna spacing at both the BS and the RIS is set to half the wavelength, i.e., $d_{\mathrm{BS}} = d_{\mathrm{RIS}} = \lambda_c / 2$. The channel gains are modeled as $\alpha_{l} \sim \mathcal{CN}(0, 10^{-3} d_{\text{BR}}^{-2.2})$ and $\beta_{k,j} \sim \mathcal{CN}(0, 10^{-3} d_{\text{RU}}^{-2.8})$, where $d_{\text{BR}}$ and $d_{\text{RU}}$ denote the distances between the BS and RIS, and between the RIS and users, respectively. The number of users is $K=5$. The RIS-BS link has $L=4$ paths, and each user-RIS link has $J_1 = J_2 = \cdots = J_K = 3$ paths. The SNR is defined as $10 \log\left(10^{-6} d_{\mathrm{BR}}^{-2.2} d_{\mathrm{RU}}^{-2.8} p / \delta^{2}\right)$, with the per-user transmit power set to $p = 1 \, \text{W}$. The normalized mean square error (NMSE) serves as the performance metric, defined as
$$
\text{NMSE} = \mathbb{E}\left[ \frac{\sum_{k=1}^{K} \| \widehat{\mathbf{G}}_{k} - \mathbf{G}_{k} \|_{F}^{2}}{\sum_{k=1}^{K} \| \mathbf{G}_{k} \|_{F}^{2}} \right],
$$
where $\widehat{\mathbf{G}}_{k}$ is the estimated cascaded channel matrix.

We compare the proposed three-stage channel estimation protocol with two algorithms in BD-RIS scenario. The features of each algorithm are as follows:
\begin{itemize}
    \item \textbf{Proposed protocol}: The three-stage channel estimation protocol developed in this work exploits the \textit{group-connected} block-diagonal BD-RIS structure alongside the inherent angular-domain sparsity of channels. To demonstrate the general applicability of our protocol across varying conditions, we evaluate the proposed protocol under two distinct number of groups, $G=4$ and $G=9$.
    \item \textbf{Direct-OMP}: This algorithm formulates the estimation problem through the vectorization of the cascaded received signal: $\mathrm{vec}(\mathbf{Y}_{k})=\sqrt{p}\left( \boldsymbol{\Theta}_{k}^{\mathrm{T}} \otimes  \mathbf{I}_{N}\right)\mathrm{vec}(\mathbf{G}_{k})+\mathrm{vec}(\mathbf{N}_{k})$, inherited from Eq.~$\eqref{eq:stacked_measurement}$. It consequently attempts to solve a massive $L^{2}J_{k}$-dimensional sparse recovery problem simultaneously. Such a straightforward multi-dimensional approach inherently ignores the composite array structures, incurring prohibitive computational complexity and demanding an unsustainable pilot overhead to guarantee successful grid matching.
    \item \textbf{SBL}: As a probabilistic algorithm, the Sparse Bayesian Learning (SBL) estimator \cite{zyang2013data} is applied to recover the channel components via expectation-maximization (EM) iterations. Following the removal of the common AoA components, the equivalent measurement matrix $\tilde{\mathbf{Y}}_{k}$ outlined in Eq.~$\eqref{eq:esttrans}$ is processed by SBL to infer the posterior distributions. In our simulations, this iterative process is configured with a maximum limit of $50$ iterations and a convergence threshold of $10^{-6}$.
\end{itemize}

\begin{figure}[t]
    \centering
    \includegraphics[width=0.41\textwidth]{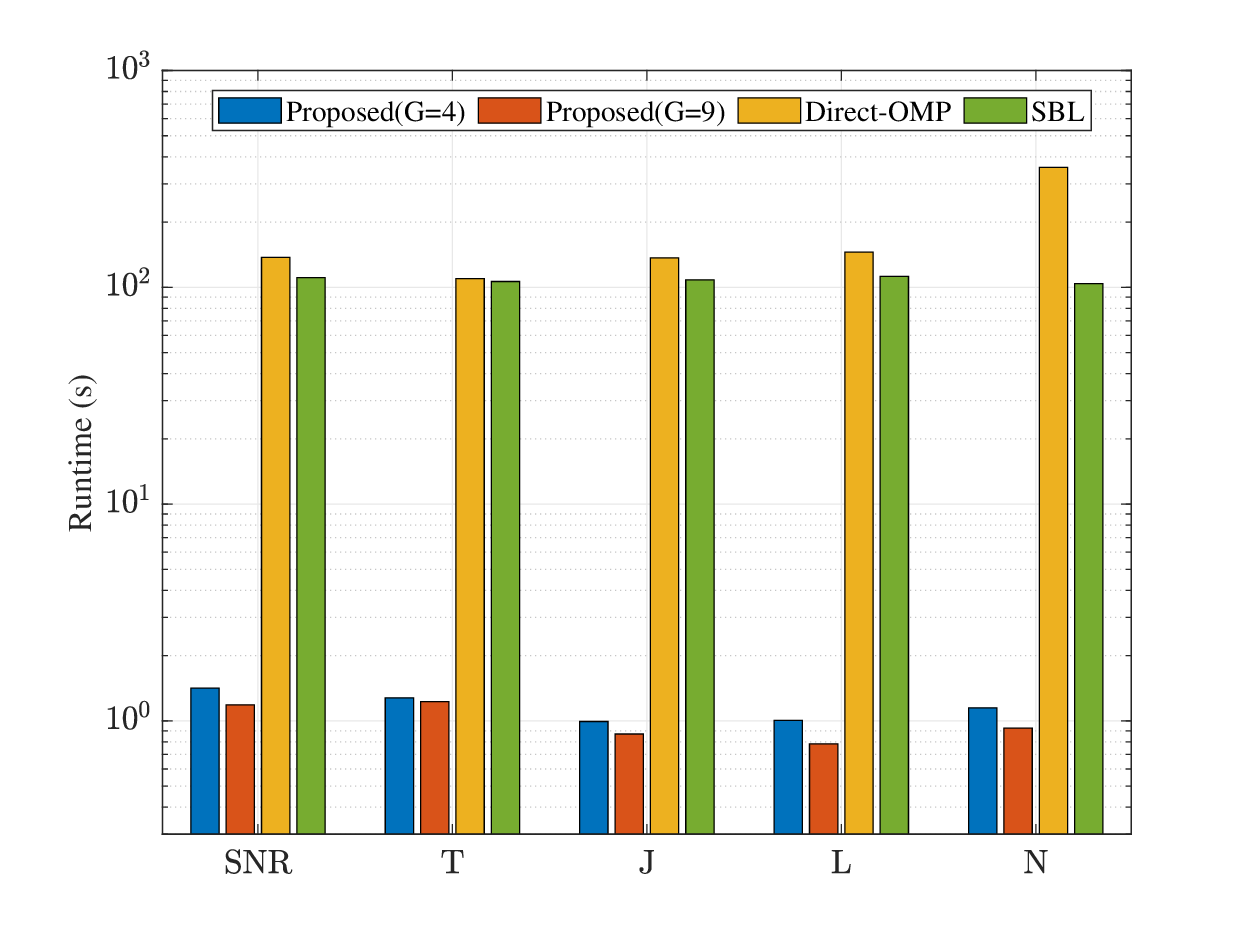}
    \caption{Runtime of algorithms with different parameters. }
    \label{fig:cpx}
\end{figure}
To validate the computational efficiency in Section~\ref{sec:pilot_and_complexity}, Fig.~\ref{fig:cpx} illustrates the average runtime required by each algorithm under various parameters. Consistent with our theoretical complexity analysis, the proposed three-stage protocol exhibits a significantly lower computational burden compared to the baseline alternatives. By estimating the cascaded channel in sequential stages, our approach effectively avoids the extremely large search grid that affects the conventional Direct-OMP and iterative SBL algorithms.

\begin{figure}[t]
    \centering
    \includegraphics[width=0.41\textwidth]{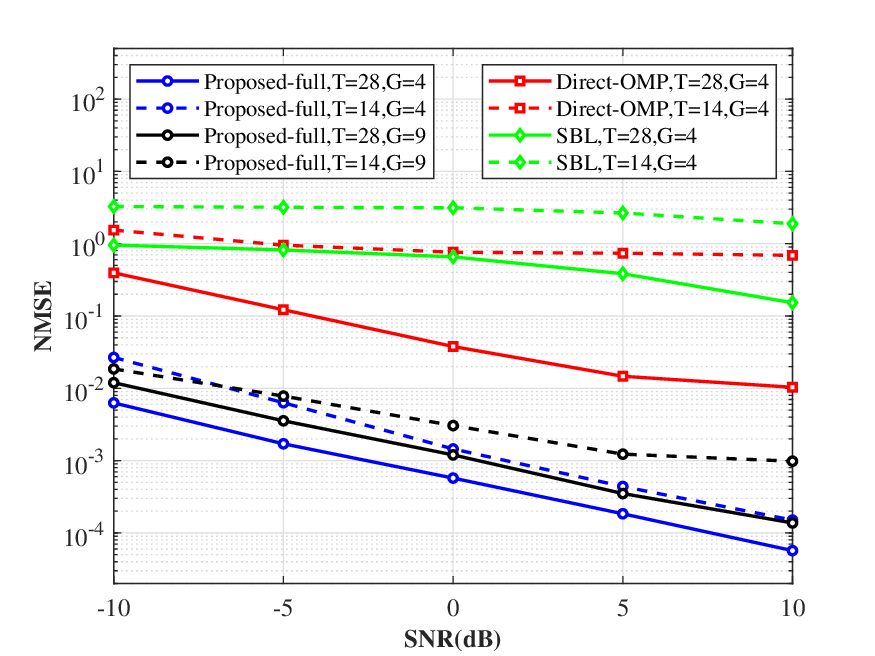}
    \caption{NMSE versus SNR.}
    \vspace{-2pt}
    \label{fig:nmse_vs_snr}
\end{figure}

Fig.~\ref{fig:nmse_vs_snr} presents the NMSE performance of the various channel estimation algorithms against the SNR. We evaluate two average pilot overhead budgets: $\mathrm{T}=28$ ($\tau_1 = 48$ for the typical user and $\tau_k = 24$ for others) and a reduced budget of $\mathrm{T}=14$ ($\tau_1 = 29$ and $\tau_k = 11$). This asymmetric allocation prioritizes the typical user to ensure reliable extraction of common parameters in Stages I and II, thereby mitigating error propagation in Stage III. The curves demonstrate that the proposed protocol achieves a significant reduction in NMSE as the SNR increases, with the $G=4$ setup outperforming $G=9$. This performance gap arises because a larger $G$ degrades the column orthogonality of the dictionary, which negatively affects the estimation accuracy. In contrast, the Direct-OMP algorithm exhibits inferior performance under $\mathrm{T}=28$ and fails under the low $\mathrm{T}=14$ budget due to grid mismatch. Across the evaluated transmit power range, the SBL method is observed to yield high NMSE. This can be attributed to the inherent difficulty of obtaining an accurate estimate for the noise covariance matrix in BD-RIS-aided systems. For a standardized comparison, a reference SNR of $0$~dB is adopted for all subsequent simulations.

\begin{figure}[t]
    \centering
    \includegraphics[width=0.41\textwidth]{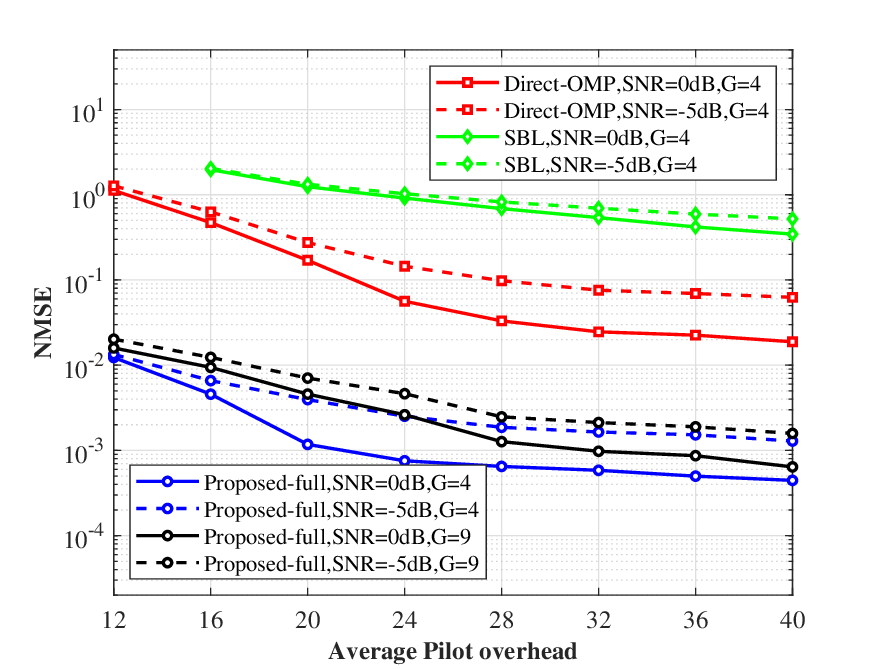}
    \caption{NMSE versus pilot overhead. }
    \label{fig:nmse_vs_pilot}
\end{figure}

The impact of the average number of pilot symbols $\mathrm{T}$ on the NMSE is depicted in Fig.~\ref{fig:nmse_vs_pilot}, evaluated at SNR regimes of $0$~dB and $-5$~dB. The proposed protocol consistently outperforms the baseline algorithms across the entire range of $\mathrm{T}$. In severely pilot-constrained regions, the Direct-OMP algorithm fails to converge because its large-scale sensing matrix requires massive observations to fulfill the restricted isometry property (RIP) \cite{candes2006}. As $\mathrm{T}$ increases, the NMSE decreases monotonically for all methods due to the larger number of observations. However, the NMSE of the proposed protocol decreases more rapidly, while the SBL-based algorithm saturates with a relatively high error floor. The performance improvement of the proposed protocol in the low overhead regime highlights its advantage. Additionally, the recurring trend where $G=9$ exhibits higher estimation errors than $G=4$ further verifies that a larger group size reduces dictionary orthogonality.

\begin{figure}[t]
    \centering
    \includegraphics[width=0.41\textwidth]{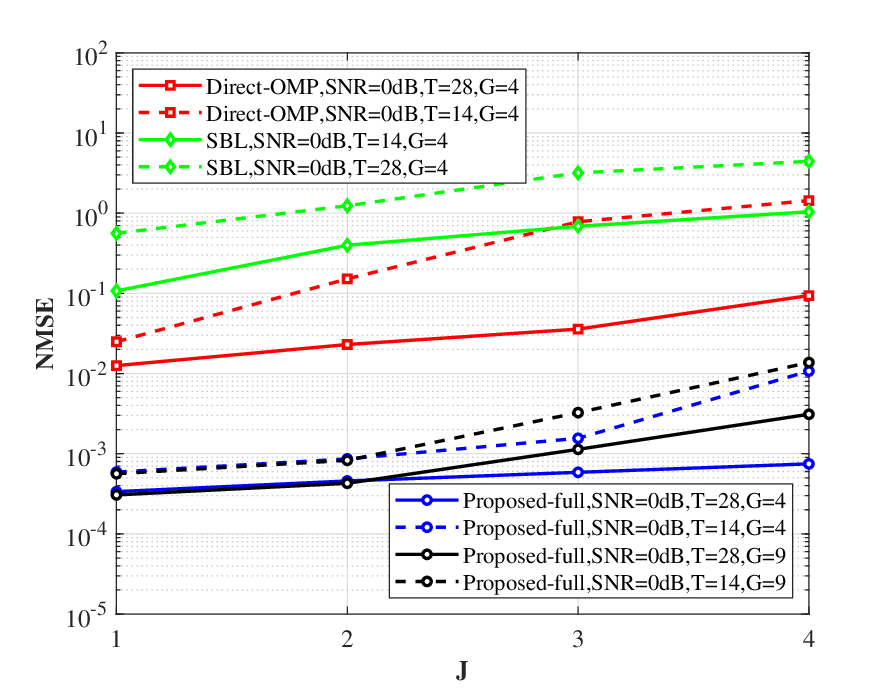}
    \caption{NMSE versus the number of user-RIS link $J$ with $\text{SNR} =0$~dB.}      
    \label{fig:nmse_vs_J}
\end{figure}

Fig.~\ref{fig:nmse_vs_J} characterizes the impact of the number of paths $J$ in the user-RIS link, on the NMSE. While an overall performance degradation affects all estimators as $J$ increases, the proposed protocols maintain a significant accuracy advantage over both Direct-OMP and SBL-based algorithms. This performance degradation occurs because a larger $J$ increases the total number of unknown angular and path gain parameters, making the estimation problem more complex. Additionally, identifying the multi-path support set becomes more challenging. Nevertheless, the proposed configuration with $G=4$ under the $\mathrm{T}=28$ budget still maintains robust performance, showing the advantage of utilizing the grouped BD-RIS structure in environments with rich multipath components.

\begin{figure}[t]
    \centering
    \includegraphics[width=0.41\textwidth]{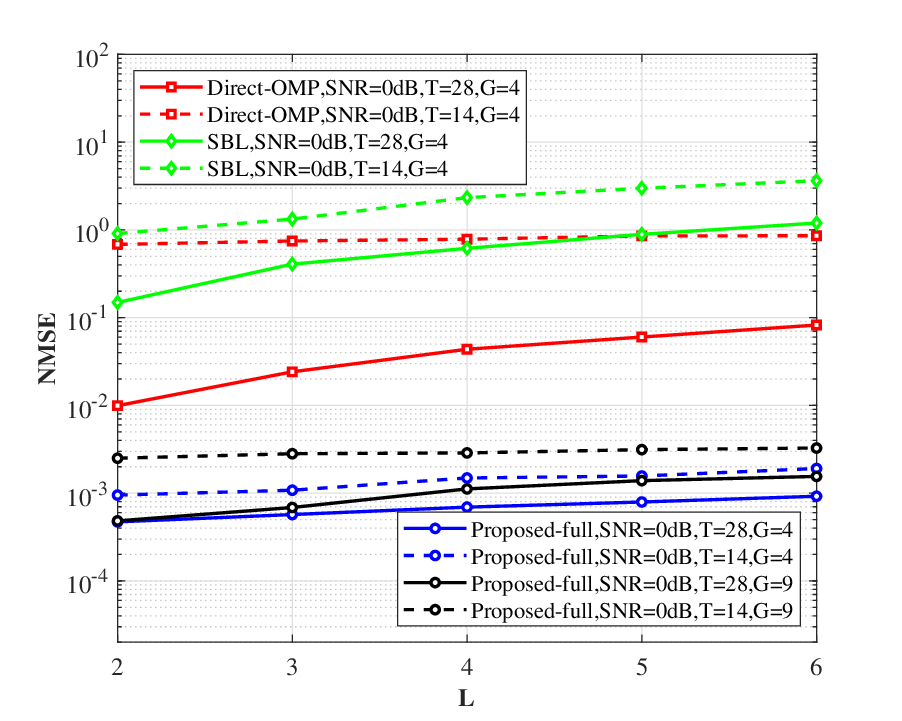}
    \caption{NMSE versus the number of paths $L$ with $\text{SNR} =0$~dB.}      
    \label{fig:nmse_vs_L}
\end{figure}

Similarly, Fig.~\ref{fig:nmse_vs_L} visualizes the NMSE trend as a function of $L$, the number of propagation paths in the common RIS-BS link. Although all estimators experience a steady rise in NMSE as $L$ scales up, the proposed protocol preserves its clear superiority over the benchmarks across all parameter configurations. Moreover, compared to the impact of $J$, the performance degradation caused by the increase of $L$ is relatively smaller. This is due to the different roles these parameters play in the structural block-Kronecker layout: an increase in $J$ directly scales up the sparsity level of the individual user-RIS channels to be recovered, whereas $L$ mainly expands the dimension of the common parameter matrix. By sequentially estimating these components, the proposed protocol effectively handles this parameter expansion and achieves reliable estimation.
\begin{figure}[t]
    \centering
    \includegraphics[width=0.41\textwidth]{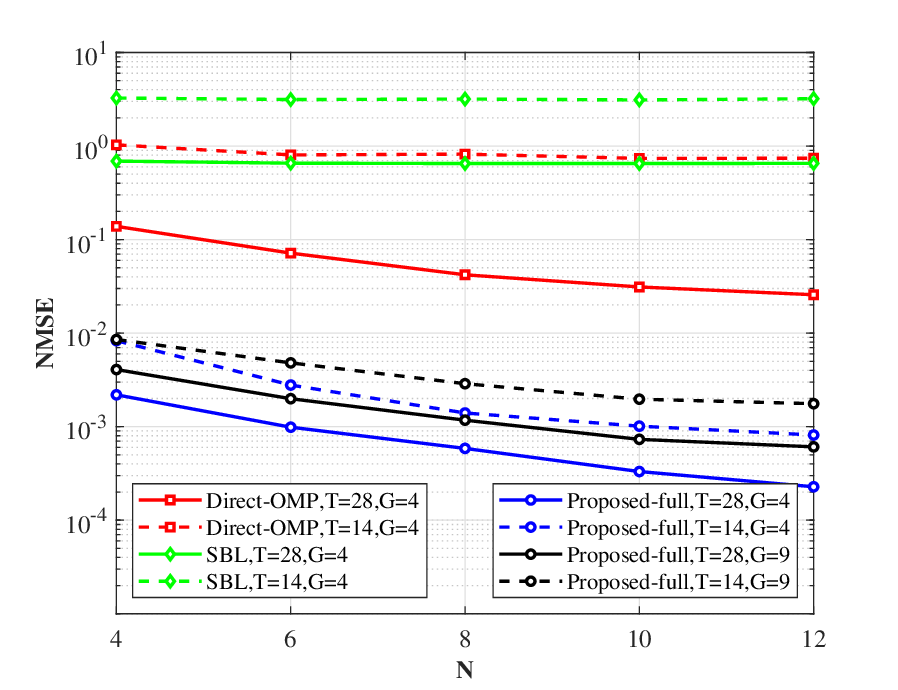}
    \caption{NMSE versus the number of BS antennas $N$: $N=N_v\times N_h,N_v=N_h$, with $\text{SNR} =0$~dB.}    \label{fig:nmse_vs_N}
\end{figure}

Finally, Fig.~\ref{fig:nmse_vs_N} illustrates the NMSE versus the number of BS antennas $N$. As $N$ increases, the larger array aperture tightens the beamwidth and sharpens the spatial resolution, which naturally improves the channel estimation precision across all evaluated algorithms. The proposed protocol consistently achieves the lowest NMSE, and its performance advantage over the baselines widens as $N$ grows. This gain is mainly because a larger receiving array provides more spatial DoF, which can be effectively utilized by the DFT-based peak extraction in Stage I. Consequently, while the Direct-OMP and SBL algorithms yield poor estimation accuracy with the high-dimensional noise, our proposed protocol maintains a steady improvement in accuracy. This demonstrates the practical suitability of the proposed protocol for massive MIMO networks with large-scale antenna arrays.

\section{Conclusion}
\label{subsec:conclusion}
This paper proposed a three-stage channel estimation protocol for BD-RIS-aided MU millimeter-wave systems. The protocol effectively exploits the block-diagonal structure of group-connected BD-RIS and the cascaded channel sparsity. 
In Stage I, the common AoAs at the BS are estimated using a DFT-based method combined with angle rotation. 
This provides the common angular parameters for subsequent stages and reduces the pilot overhead for common parameters. 
Stage II acquires the full CSI for a typical user. 
By exploiting the linear relationships between paths via an OMP-based approach, this stage maintains estimation accuracy while significantly lowering computational complexity. 
Stage III estimates the channels for the remaining users with reduced overhead, 
accomplished by the proposed HBOMP algorithm. 
The HBOMP leverages the structural prior of the common RIS-BS channel obtained from the typical user, 
thereby further minimizing the pilot and computational costs for multi-user estimation. 
The proposed protocol achieves superior estimation accuracy, higher pilot efficiency, and lower com putational complexity compared to existing algorithms, as validated by simulation results.

\end{document}